\documentclass[english,journal=jctcce,manuscript=article,email=true,etalmode=truncate,maxauthors=0]{achemso}

\usepackage{graphicx}
\graphicspath{{figures/}}
\usepackage{physics}
\usepackage{multirow}
\usepackage{xcolor}
\usepackage{caption}
\usepackage{subcaption}
\usepackage{bm}

  \usepackage[numbers,sort&compress,super]{natbib}
  \usepackage{natmove}

\usepackage{hyperref} 
\hypersetup{colorlinks=true, citecolor=blue, urlcolor=blue, linkcolor=blue}
\usepackage{cleveref}	
  	\crefname{figure}{Figure}{Figures}
  	\crefname{table}{Table}{Tables}
  	\crefname{equation}{Eq.}{Eqs.}
  	\crefname{section}{Section}{Sections}
  	\crefname{subsection}{Section}{Sections}
  	\crefname{subsubsection}{Section}{Sections}
  	\crefname{algorithm}{Algorithm}{Algorithms}
  	
\newcommand{\wann}[2]{\tilde{\psi}_{#1,#2}}
\newcommand{\co}[2]{\psi_{#1,#2}}
\newcommand{\bloch}[2]{\phi_{#1,#2}}
\newcommand{\AO}[2]{\phi_{#1,#2}}

\newcommand{\code}[1]{\texttt{#1}}

\usepackage{todonotes}

\newcommand{\redBox}[1]{\fcolorbox{red}{white}{#1}}
\newcommand{\blueBox}[1]{\fcolorbox{blue}{white}{#1}}
\newcommand{\orangeBox}[1]{\fcolorbox{orange}{white}{#1}}

\newcommand{\greenBox}[1]{\fcolorbox{green}{white}{#1}}
\title{Robust Pipek-Mezey Orbital Localization in Periodic Solids}

\author{Marjory C. Clement}

\author{Xiao Wang}
\altaffiliation{Center for Computational Quantum Physics, Flatiron Institute, New York, New York 10010, USA}

\author{Edward F. Valeev}
\affiliation{Department of Chemistry, Virginia Tech, Blacksburg, VA 24061}
\email{efv@vt.edu}

\begin{document}

\date{\today}

\begin{abstract}
We describe a robust method for determining Pipek-Mezey (PM) Wannier functions (WF), recently introduced by J\'onsson et al. ({\em J. Chem. Theor. Chem.} {\bf 2017}, 13, 460), which provide some formal advantages over the more common Boys (also known as maximally-localized) Wannier functions. The Broyden-Fletcher-Goldfarb-Shanno (BFGS) based PMWF solver is demonstrated to yield dramatically faster convergence compared to the alternatives (steepest ascent and conjugate gradient) in a variety of 1-, 2-, and 3-dimensional solids (including some with vanishing gaps), and can be used to obtain Wannier functions robustly in supercells with thousands of atoms. Evaluation of the PM functional and its gradient in periodic LCAO representation used a particularly simple definition of atomic charges obtained by Moore-Penrose pseudoinverse projection onto the minimal atomic orbital basis. An automated ``Canonicalize Phase then Randomize'' (CPR) method for generating the initial guess for WFs contributes significantly to the robustness of the solver.
\end{abstract}

\maketitle

\section{Introduction}
Localized orbitals, by eliminating the artifacts of symmetry and accidental degeneracy, are valuable for qualitative interpretation of electronic states in terms of traditional concepts of chemical bonding and as a computational basis underpinning many reduced complexity algorithms in electronic structure. Localized orbitals are particularly relevant for periodic solids, where due to the lattice symmetry, the eigenstates of observables are delocalized over the entire lattice; such delocalization becomes counterproductive when the unit cell size significantly  exceeds the lengthscale of the decay of the 1-particle reduced density matrix or when the focus is on localized features of the electronic structure (e.g., impurities and surface adsorbates).

For molecules, the most commonly used {\em black-box} localization methods are those due to Foster and Boys (FB) \cite{VRG:boys:1960:RMP,FB} and Pipek and Mezey (PM);\cite{VRG:pipek:1989:JCP} several historically-important methods like Edmiston-Ruedenberg (ER)\cite{ER} and von Niessen (vN)\cite{VRG:vonniessen:1972:JCP} are rarely used nowadays. All of these orbitals are defined as the stationary points of the corresponding functional. For example, the FB orbitals minimize the sum of squared position uncertainties of the orbitals. Pipek-Mezey orbitals maximize the sum of squares of atomic charges of each orbital; the original PM definition utilized Mulliken charges, which are meaningless for non-minimal basis sets, and other definitions of atomic charges are far more robust.\cite{VRG:cioslowski:1991:JMC,VRG:alcoba:2006:JCC,VRG:knizia:2013:JCTC,VRG:lehtola:2014:JCTC,VRG:janowski:2014:JCTC} Unlike the FB orbitals and like the ER and vN orbitals, the PM orbitals preserve the $\sigma-\pi$ separation, which is an important advantage when interpreting the electronic structure.\footnote{However, unlike the ER and vN orbitals, the PM orbitals lack the intra-atomic localizaton.} Further enhancements of these functionals include the use of higher-than-second power analogs of the FB\cite{VRG:hoyvik:2012:JCP} and PM\cite{VRG:knizia:2013:JCTC,VRG:lehtola:2014:JCTC} functionals.

In periodic solids, the analogs of molecular localized orbitals are referred to as {\em generalized} Wannier functions (WF),\cite{VRG:marzari:1997:PRB} which, in contrast to conventional Wannier functions\cite{VRG:wannier:1937:PR} obtained from a single band, mix Bloch orbitals from several bands.  \cite{VRG:marzari:2012:RMP}
Marzari and Vanderbilt championed the use of the FB functional to determine the generalized Wannier functions, which they dubbed as {\em maximally-localized} (generalized) Wannier functions (MLWF).\cite{VRG:marzari:1997:PRB} Since the implementation of an MLWF solver in the \code{Wannier90} package\cite{VRG:pizzi:2020:JPCM} the use of MLWF has become popular\cite{VRG:marzari:2012:RMP} for interpreting the electronic states and as a basis in a number of reduced-scaling many-body electronic structure methods.\cite{VRG:maschio:2007:PRB,VRG:usvyat:2011:JCP}Note that the MLWF formalism is not unique;
for example, the FB localization of periodic orbitals by Zicovich-Wilson et al.,\cite{VRG:zicovich-wilson:2001:JCP} as implemented in the \code{Crystal} package,\cite{VRG:dovesi:2018:WIRCMS}
in contrast to the approach of Marzari and Vanderbilt, uses a purely real-space formulation. Furthermore, the choice of localization functional itself is not unique; MLWFs are just one of many plausible types of ``maximally-localized'' generalized Wannier functions.

Recently, J\'onsson et al. introduced the use of Pipek-Mezey Wannier functions
(PMWF),\cite{VRG:jonsson:2017:JCTC} which, in contrast to the FB-based WFs of Refs.
\citenum{VRG:marzari:1997:PRB} and \citenum{VRG:zicovich-wilson:2001:JCP} and
analogously to the molecular PM counterparts, do not mix the $\sigma$ and $\pi$
orbitals.
J\'onsson et al. evaluated PMWFs using charges defined via real-space partitioning; the atomic charges and PMWFs were found to be insensitive to the specifics of the atomic partitioning, just as molecular PM orbitals\cite{VRG:lehtola:2014:JCTC} were found to depend weakly on the choice of atomic charges
used in the PM functional.
The maximum of the PM functional was located by the conjugate gradient (CG) method, which for PM optimization in molecules was found to be superior than
the steepest ascent (SA) method.\cite{VRG:lehtola:2013:JCTC}
Unfortunately, the convergence of the CG solver for PMWFs,  as documented by J\'onsson
et al. as well as revealed in our own experiments, can be fairly poor, requiring
hundreds or even thousands of iterations. Also note that the CG is used to construct MLWFs in \code{Wannier90}.

As we discovered in our work, the performance of the SA and (nonlinear) CG solvers for
PMWFs can be greatly improved by the Broyden-Fletcher-Goldfarb-Shanno
(BFGS) solver.\cite{VRG:broyden:1970:IJAM,VRG:fletcher:1970:CJ,VRG:goldfarb:1970:MC,VRG:shanno:1970:MC} While BFGS\cite{VRG:kari:1984:IJQC} as well as other quasi-Newton
methods\cite{VRG:leonard:1982:TCA,VRG:hoyvik:2012:JCTC} have been used to solve for localized orbitals in
molecules, to our knowledge its use for Wannier function optimization has not been
considered.
Thus the main purpose of this manuscript is to document the implementation of the BFGS PMWF solver and compare its performance to that of SA and CG. Here we also document the particularly simple definition of atomic charges that we devised via pseudoinverse projection on the minimal basis, which is very convenient when employing the linear combination of atomic orbitals (LCAO) representation of the periodic solid's orbitals.
The result of this work is a robust
periodic localizer that has been successfully applied to one-, two-, and
three-dimensional systems with large Born-von-K\'arm\'an unit cells, including some with vanishingly small band gaps.

\section{Formalism}

\subsection{Basic Definitions}
The objective of PM localization is to convert an input set of
periodic (Bloch) orbitals (usually, Hartree-Fock or Kohn-Sham orbitals)
to a set of localized orbitals.
In our work, the orbitals are expressed in the LCAO representation, expanded in (contracted) Gaussian AOs, obtained by the reduced-scaling Hartree-Fock method recently
reported by some of us.\cite{VRG:wang:2020:JCP}
The $j$th Bloch orbital $\ket{\co{j}{\mathbf{k}}}$ with crystal momentum wave vector \textbf{k}
is a linear combination of Bloch AOs $\{ \bloch{\nu}{\mathbf{k}} \}$:
\begin{equation}
\label{eq:bo}
\ket{\co{j}{\mathbf{k}}}
=
\sum_{\nu}
C^{j}_{\nu,\mathbf{k}}
\ket{\bloch{\nu}{\mathbf{k}}} .
\end{equation}
Bloch AOs in turn are translation-symmetry-adapted linear combinations of AOs
 $\{\AO{\nu}{\mathbf{R}}\}$:
\begin{equation}
\label{eq:bao}
\ket{\bloch{\nu}{\mathbf{k}}}
=
\frac{1}{\sqrt{N}}
\sum_{\mathbf{R}}
e^{\mathrm{i}\mathbf{k}\cdot\mathbf{R}}
\ket{\AO{\nu}{\mathbf{R}}},
\end{equation}
with $\mathbf{R}$ denoting the origin of each primitive unit cell in the Born-von-K\'arm\'an (BvK) unit cell (``supercell'') composed of $N$ primitive cells; by convention, $\mathbf{R}=0$ corresponds to the {\em reference} unit cell. Please note the different normalization of Bloch AOs in \cref{eq:bao} than is traditional in the periodic LCAO literature.\cite{VRG:harris:1975:Tc,VRG:pisani:1980:IJQC,VRG:wang:2020:JCP}

A generalized Wannier function centered
in the unit cell at $\mathbf{R}$ will be expressed as a linear combination of Bloch orbitals:
\begin{equation}
\label{eq:WF_def}
\ket{\wann{i}{\bf{R}}}
= 
\frac{1}{\sqrt{N_{k}}}
\sum_{j,\mathbf{k}}
e^{-\mathrm{i}\mathbf{k}\cdot\mathbf{R}}
\ket{\co{j}{\mathbf{k}}}
U_{j,\mathbf{k}}^{i},
\end{equation}
where $i$ indexes the Wannier functions (localizing $o$ Bloch orbitals will produce $o$ WFs) and $N_k$ is the number of \textbf{k} points in the uniform (Monkhorst-Pack)\cite{VRG:monkhorst:1976:PRB} quadrature used to integrate the first Brillouin zone corresponding to the supercell in \cref{eq:bao} (in this work we impose $N_k = N$).
Matrices $\{\mathbf{U}_{\mathbf{k}}\}$, defined as $(\mathbf{U}_{\mathbf{k}})^i_j \equiv U^i_{j,\mathbf{k}}$,\footnote{The superscript and subscript indices on matrix elements refer to the columns and rows, respectively.} are unitary to ensure mutual orthonormality
of WFs, both associated with the same unit cell and between WFs associated with different unit cells; this also ensures that WFs span the same space as the Bloch orbitals. No additional constraints, such as realness of the WFs, are imposed.

\subsection{PM Functional with Pseudoinverse Minimal Basis Atomic Charges}
\label{sec:pseudoinverseprojection}
The PM WFs are stationary points of the PM functional:
\begin{equation}
P \equiv
\sum_{\bf{R}}
\sum_{A}
\sum_{i}
|Q^{A_{\bf{R}}}_{i}|^{p}
\label{P}
\end{equation}
where $A$ indexes the atoms in  the primitive unit cell, and $i$ indexes the Wannier functions. The atomic charge contribution
$Q^{A_{\bf{R}}}_i$ is the charge associated with orbital $i$ on atom $A$
in unit cell $\mathbf{R}$, and the charge exponent, $p$, is equal to 2 (conventional PM functional) or 4 (fourth-order PM functional\cite{VRG:knizia:2013:JCTC,VRG:lehtola:2014:JCTC}).

The original work by Pipek and Mezey used standard Mulliken charges,\cite{VRG:pipek:1989:JCP}
\begin{align}
\label{eq:q_mulliken}
Q_{i}^{A_\mathbf{R}} = \bra{\wann{i}{\bf{0}}} \hat{P}_{A_\mathbf{R}} \ket{\wann{i}{\bf{0}}},
\end{align}
where
\begin{align}
\label{eq:P_A}
\hat{P}_{A_\mathbf{R}} \equiv \sum_{\mu \in A}
\Big|\phi_{\mu,\mathbf{R}}\Big\rangle
\bra{\hat{\phi}_{\mu,\mathbf{R}}}
\end{align}
is the projector onto the AOs centered on atom $A_\mathbf{R}$ defined via the biorthogonal AO basis as
\begin{align}
\bra{\hat{\phi}_{\mu,\mathbf{R}}} \equiv & \sum_{\nu,\mathbf{R}'} (\mathbf{S}^{-1})^{\nu,\mathbf{R}'}_{\mu,\mathbf{R}} \bra{\phi_{\nu,\mathbf{R}'}} ,
\end{align}
or its Bloch-AO equivalent,
\begin{align}
\bra{\hat{\phi}_{\mu,\mathbf{R}}} \equiv & \sum_{\nu,\mathbf{k}} (\mathbf{S}^{-1})^{\nu,\mathbf{k}}_{\mu,\mathbf{R}} \bra{\phi_{\nu,\mathbf{k}}},
\end{align}
and $\mathbf{S}^{-1}$ denotes the inverse overlap matrix.

The key issue with the Mulliken charges as defined by \cref{eq:q_mulliken} is their ill-defined nature for non-minimal basis sets. This impacts the robustness of the PM localization method by making the PM orbitals sensitive to variations in the orbital basis set (OBS) and geometry. Luckily, as noted by many\cite{VRG:cioslowski:1991:JMC,VRG:alcoba:2006:JCC,VRG:knizia:2013:JCTC,VRG:lehtola:2014:JCTC,VRG:janowski:2014:JCTC}, alternative charge definitions make the PM functional much more robust (the interpretation of approaches in Refs. \citenum{VRG:cioslowski:1991:JMC,VRG:alcoba:2006:JCC} as PM with non-Mulliken charge definitions was pointed out in Ref. \citenum{VRG:lehtola:2014:JCTC}).
To this end, J\'onsson et al. utilized real-space partitioning of orbital charge densities to obtain robust atomic charges for use in the PM functional for periodic solids. \cite{VRG:jonsson:2017:JCTC}

In this work, we defined the PM functional using atomic charges evaluated with the help of a pre-defined minimal basis set (MBS) of atomic orbitals. 
Such  projection onto an MBS has been long employed to define OBS-independent atomic partitioning of densities and thus, due to its basis independent (intrinsic) nature, eliminates the undue sensitivity of charges to the orbital basis. The use of such MBS-projected charges, in the context of PM localization, has been frequently employed recently\cite{VRG:west:2013:JCP,VRG:knizia:2013:JCTC,VRG:lehtola:2014:JCTC,VRG:janowski:2014:JCTC} but the idea is much older: already Mulliken remarked that, ``{\em The ideal LCAO-MO population analysis would perhaps be in terms of free atom SCF AO's}''\cite{VRG:mulliken:1962:JCP} and goes back even earlier.

There is no unique method to use an MBS to define atomic charges, due to the arbitrariness of how to partition orbitals and/or their charge densities into atomic components. It is also far too easy to
break desired invariants, such as the equality of the sum of atomic populations and the number of electrons. For example, replacing OBS AO $\phi_{\mu,\mathbf{R}}$ by MBS AO $\bar{\phi}_{\mu,\mathbf{R}}$ in \cref{eq:P_A} (such replacement is only meaningful if the overlap inverse is defined, i.e., if the MBS is a subset of the OBS) violates such an invariant. Thus, existing approaches\cite{VRG:lu:2004:JCP,VRG:laikov:2011:IJQC,VRG:west:2013:JCP,VRG:knizia:2013:JCTC,VRG:janowski:2014:JCTC} utilize the MBS indirectly, as a way to define an MBS-like subspace of the OBS. Here we use a simpler approach.
Please note that, throughout the following discussion, quantities with an overbar
are expressed in the MBS AO representation.

Consider the OBS AO representation of the
Bloch orbitals, obtained by plugging \cref{eq:bao} into \cref{eq:bo}:
\begin{align}
    \ket{\co{i}{\bf{k}}} = \sum_{\mu,\mathbf{R}} C^{i,\mathbf{k}}_{\mu,\mathbf{R}} \ket{\phi_{\mu,\mathbf{R}}},
\end{align}
with $C^{i,\mathbf{k}}_{\mu,\mathbf{R}}$ defined as:
\begin{align}
\label{eq:CikmuR}
    C^{i,\mathbf{k}}_{\mu,\mathbf{R}} \equiv \frac{1}{\sqrt{N}} C^{i}_{\mu,\mathbf{k}} e^{\mathrm{i}\mathbf{k}\cdot\mathbf{R}};
\end{align}
clearly, $C^{j,\mathbf{k}}_{\nu,\mathbf{R}} = C^{j,\mathbf{k}}_{\nu,\mathbf{0}} e^{\mathrm{i}\mathbf{k}\cdot\mathbf{R}}$.
{\em Biorthogonal} mapping of Bloch functions on the MBS,
\begin{align}
    \ket{\overline{\co{i}{\bf{k}}}} = \sum_{\mu,\mathbf{R}} \overline{C}^{i,\mathbf{k}}_{\mu,\mathbf{R}} \ket{\bar{\phi}_{\mu,\mathbf{R}}},
\end{align}
is trivially obtained by solving
\begin{align}
\label{eq:Cmbs}
    \delta_{ij} = \bra{\co{i}{\bf{k}}} \ket{\overline{\co{j}{\bf{k}}}} = \sum_{\mu,\mathbf{R}} \bra{\co{j}{\bf{k}}} \ket{\bar{\phi}_{\mu,\mathbf{R}}} \overline{C}^{j,\mathbf{k}}_{\mu,\mathbf{R}}
\end{align}
for coefficients $\overline{C}^{j,\mathbf{k}}_{\mu,\mathbf{R}}$. A least-squares solution is produced by the Moore-Penrose pseudoinverse of the overlap matrix  between the target set of Bloch orbitals for the given $\mathbf{k}$ with the MBS AOs in the reference cell, $\bar{\phi}_{\mu,\mathbf{0}}$:
\begin{align}
\label{eq:Smbsco}
({\bf S})^{\mu}_{i,\mathbf{k}} \equiv \bra{\co{i}{\bf{k}}}\ket{\bar{\phi}_{\mu,\mathbf{0}}}
   = \sum_{\nu,\mathbf{R}} \left(C^{i,\mathbf{k}}_{\nu,\mathbf{R}}\right)^* \bra{\phi_{\nu,\mathbf{R}}}\ket{\bar{\phi}_{\mu,\mathbf{0}}} .
\end{align}
Only simple Gaussian AO overlaps are needed to evaluate \cref{eq:Smbsco}, with the lattice sum in \cref{eq:Smbsco} geometrically convergent. Evaluation in other numerical representations, such as plane waves (PWs), should be also straightforward.
Extension of such \emph{pseudoinverse} MBS mapping to the molecular case is obvious, where such procedure is related to how the {\em corresponding} orbitals are constructed.\cite{VRG:king:1967:JCP} Note also that the pseudoinverse-mapped orbitals are not orthonormal, hence the pseudoinverse charges differ from the existing definitions for minimal-basis-derived charges;\cite{VRG:lu:2004:JCP,VRG:laikov:2011:IJQC,VRG:west:2013:JCP,VRG:knizia:2013:JCTC,VRG:janowski:2014:JCTC} however the relationship between the pseudoinverse MBS charges and other MBS-based charges is outside of the scope of this article and will be discussed elsewhere.

The OBS AO coefficients of the WF,
\begin{align}
    W^{i}_{\mu,\mathbf{R}} \equiv \frac{1}{\sqrt{N_{k}}}
\sum_{j,\mathbf{k}}
C^{j,\mathbf{k}}_{\mu,\mathbf{R}}
U_{j,\mathbf{k}}^{i},
\end{align}
are mapped straightforwardly to the MBS AO coefficients:
\begin{align}
    \overline{W}^{i}_{\mu,\mathbf{R}} \equiv \frac{1}{\sqrt{N_{k}}}
\sum_{j,\mathbf{k}}
\overline{C}^{j,\mathbf{k}}_{\mu,\mathbf{R}}
U_{j,\mathbf{k}}^{i}.
\end{align}
The minimal-basis pseudoinverse charges are thus obtained by replacing the OBS with the MBS in \cref{eq:q_mulliken} and replacing the corresponding OBS AO coefficient of the WF,  $\bra{\bar{\phi}_{\mu,\mathbf{R}}}\ket{\wann{i}{\bf{0}}} \equiv W^{i}_{\mu,\mathbf{R}} $, with the corresponding MBS WF AO coefficient:
\begin{align}
    \label{eq:q_pseudo}
\overline{Q}_{i}^{A_\mathbf{R}} = \frac{1}{2} \left[ \sum_{\mu \in A}  \bra{\wann{i}{\bf{0}}}\ket{\bar{\phi}_{\mu,\mathbf{R}}} \overline{W}^{i}_{\mu,\mathbf{R}} + \mathrm{h.c.} \right],
\end{align}
where ``h.c.'' denotes the Hermitian conjugate. Evaluation of these charges in the LCAO representation leverages the Bloch-MBS overlaps (\cref{eq:Smbsco}) and is therefore completely straightforward. For a fixed MBS these charges are expected to depend weakly on the OBS and have a well-defined basis set limit.

To avoid introducing a new symbol, $P$ will henceforth denote the PM functional (\cref{P}) defined with the MBS pseudoinverse charges.

\subsection{PM Functional Maximization}
\label{sec:solvers}

\subsubsection{Initial WF Guess}
\label{sec:initialguess}
It is easy to see that the WFs, defined by the unitary matrices $\{\mathbf{U}_{\mathbf{k}}\}$ in \cref{eq:WF_def}, are in general not uniquely defined by the corresponding functional. The nonuniqueness stems from several factors. First, PM and other functionals defining WFs are invariant with respect to arbitrary permutations of the sequence of WFs. This may appear trivial, but in general means that comparing sets of WFs produced in 2 separate computations is not straightforward (see e.g. Ref. \citenum{VRG:jonsson:2017:JCTC}). Second, the WF functional is invariant with respect to all or some of the geometric transformations of the space group of the crystal (such as shifting a WF by a lattice vector). Third, the functionals defining WFs routinely have multiple maxima for a given system; hence, finding the global maximum is NP-hard. Thus, WF computation relies on heuristics to generate initial guesses for $\{\mathbf{U}_{\mathbf{k}}\}$; this initial guess and other solver details determine which functional maximum will be located.

Initial guesses for generalized WFs are typically obtained by projecting Bloch orbitals onto some trial functions. For example, Marzari and Vanderbilt utilized Gaussians located at expected centers of charge of Wannier functions, such as midbond centers;\cite{VRG:marzari:1997:PRB}
such user-controlled guess construction is also utilized by \code{Wannier90}.\cite{VRG:pizzi:2020:JPCM}
A more automated approach was used by Zicovich-Wilson et al.,\cite{VRG:zicovich-wilson:2001:JCP} who approximately projected Bloch orbitals onto the reference cell's AO basis (since they expanded the Bloch orbitals in  an AO basis already, such projection was trivial); unfortunately, such a choice is not appropriate when covalent bonds cross boundaries of the unit cell. A similar approach was used by Mustafa et al.,\cite{VRG:mustafa:2015:PRB} who projected Bloch orbitals (expanded in PW basis) onto an appropriate set of AOs that spanned a space containing the target Wannier set. To account for the covalent bonds crossing the unit cell boundaries, the projection AO set must include AOs not only in the reference cell but also in its nearest periodic images.

Note that these projection-based approaches are still not robust enough to deal with multiple minima, since it is necessary to generate multiple initial guesses to probe the global optimality of the resulting WFs. Thus, J\'onsson et al.\cite{VRG:jonsson:2017:JCTC} simply generated guess WFs using randomly-generated $\{\mathbf{U}_{\mathbf{k}}\}$ and ran multiple computations.

Here we have devised a novel automated ``Canonicalize Phase then Randomize'' (CPR) method for generating guess WFs that can be applied to arbitrary Bloch orbitals expressed in LCAO and non-LCAO representations. The first step in this method is motivated by the realization that, to produce localized Wannier functions even for bands composed of a single atomic orbital (e.g., core bands), it is helpful to canonicalize the phases of the AO coefficients at different $\mathbf{k}$ points. Such phase canonicalization can be viewed as removing the gauge freedom of the Bloch orbitals; once the arbitrariness of the gauge is removed, then the original Wannier prescription\cite{VRG:wannier:1937:PR}, \cref{eq:WF_def} with $\{\mathbf{U}_{\mathbf{k}}\}$ set to identity, will recover the maximally-localized state, namely, the atomic orbital in the reference cell. Of course, the generalized Wannier functions can compensate for the gauge freedom of the Bloch orbitals via the $\mathbf{k}$-dependence of $\{\mathbf{U}_{\mathbf{k}}\}$; thus, the generalized Wannier orbital for a single-AO band with arbitrary gauge transformation will still be a single AO (although it will not necessarily be the reference cell AO). But by including the phase canonicalization of the input Bloch orbitals, the WF functional maximization becomes more robust by starting from a good initial guess.

Phase canonicalization in the CPR method proceeds as follows:
\begin{itemize}
\item The set of Bloch orbitals at the $\Gamma$ point ($\mathbf{k}=\mathbf{0}$) is split into degenerate subsets (bands). A set of orbitals is considered degenerate if its eigenenergies are within a prescribed energy tolerance $\epsilon$.
\item For each band $\alpha$, the {\em phase-defining} subset of AOs, $\{\mu_\alpha\}$, includes AOs $\mu$ that have the largest occupancies in the band:
\begin{align}
    \rho^{\alpha,\mathbf{0}}_{\mu,\mathbf{0}} \equiv \sum_{i_{\alpha}} |C^{i_\alpha,\mathbf{0}}_{\mu,\mathbf{0}}|^2,
\end{align}
where $i_\alpha$ are the Bloch orbitals in band $\alpha$. The phase-defining AO set, $\{\mu_\alpha\}$, thus includes the AOs with the greatest contribution to the band; it can include a single AO (most common) or multiple AOs (due to geometric symmetry and band degeneracy).
\item The phase of every Bloch orbital, $i$, at the $\Gamma$ point is aligned so that the coefficient of the {\em first} phase-defining AO, $\mu^{(0)}_{\alpha_i}$, for its band, $\alpha_i$, is positive:
\begin{align}
\label{eq:canongamma}
    C^{i}_{\mu,\mathbf{0}} \to C^{i}_{\mu,\mathbf{0}} \times \left(\frac{\left|C^{i}_{\mu^{(0)}_{\alpha_i},\mathbf{0}}\right|}{C^{i}_{\mu^{(0)}_{\alpha_i},\mathbf{0}}}\right).
\end{align}
This step eliminates possible arbitrary phase factors that were introduced by the SCF solver and makes all AO coefficients at the $\Gamma$ point real.
\item Bloch orbitals at every $\mathbf{k}$ point are next mapped to the matching orbital at the $\Gamma$ point. To perform such matching consider orbitals $i$ and $j$ at two neighboring points $\mathbf{k}$ and $\mathbf{k}'$, respectively. Their refcell MBS ``overlap'' is defined as the overlap of Bloch orbital $\co{i}{\bf{k}}$ with the Bloch orbital $\co{j}{\bf{k}'}$ pseudoinverse-projected (see \cref{sec:pseudoinverseprojection}) onto the reference cell's MBS AO basis:
\begin{align}
    \bar{S}^{i,\mathbf{k}}_{j,\mathbf{k}'} \equiv 
    \sum_\mu \bra{\co{i}{\bf{k}}}\ket{\bar{\phi}_{\mu,\mathbf{0}}} \overline{C}^{j,\mathbf{k}'}_{\mu,\mathbf{0}}.
\end{align}
All matrix elements in this equation were already evaluated when computing the MBS AO projections of the Bloch orbitals for the purpose of computing atomic charges.
For every $i\in[0,o)$, where $o$ is the number of Bloch orbitals being localized, $\co{i}{\mathbf{k}}$ {\em matches} orbital $\co{j}{\mathbf{k}'}$ point if  $|\bar{S}^{i,\mathbf{k}}_{j,\mathbf{k}'}|$ is the largest among all $j$; if the match candidate $\co{j}{\bf{k}'}$ has been declared a match for a another orbital $i'<i$ next best candidate is chosen. If orbital $\co{i}{\mathbf{k}}$ was matched to orbital $\co{j}{\mathbf{k}'}$ its phase is canonicalized such that its reference-cell MBS overlap is real:
\begin{align}
    C^{i}_{\mu,\mathbf{k}} \to C^{i}_{\mu,\mathbf{k}} \times \left(\frac{\left| \bar{S}^{i,\mathbf{k}}_{j,\mathbf{k}'} \right|}{\bar{S}^{i,\mathbf{k}}_{j,\mathbf{k}'}}\right).
\end{align}
Since the goal is align the phases of bands at all $\mathbf{k}$ points to the bands at the $\Gamma$ point the band matching is performed using sequences of $\mathbf{k}$ points originating from the $\Gamma$ point, that span the entire first Brillouin zone mesh. Assuming that points in a uniform mesh of $\mathbf{k}$ points are indexed by triplets $(i,j,k)$, with $i,j,k = 0, \pm 1, \pm 2 ... \pm \left \lfloor{N_k/2}\right \rfloor$, for 1 dimensional structures bands at the mesh points $(1, 0, 0)$ and $(-1, 0, 0)$ are matched to the canonicalized bands at (0,0,0) ($\Gamma$ point), then bands at mesh points $(2, 0, 0)$ and $(-2, 0, 0)$ are matched to the canonicalized bands at $(1,0,0)$ and $(-1,0,0)$, respectively, and so on. Similarly, for a 2 dimensional structure the bands at $(i, 1, 0)$ and $(i, -1, 0)$ are matched to the canonicalized bands at  $(i, 0, 0)$, etc.
\item To account for band crossings bands at point $\mathbf{k}$ are sorted to appear in the same order as their matching bands at point $\mathbf{k}'$.
\end{itemize}
Despite its relative simplicity, the phase canonicalization is fairly robust and significantly improves the quality of the trial Wannier functions (see Supporting Information). 
However, the described algorithm is not perfect: (1) it requires a dense Brillouin zone mesh to track high-dispersion bands across the Brillouin zone reliably, (2) it relies on an {\em ad hoc} way of matching bands, and (3) it does not account for arbitrary rotations among the degenerate bands.
Work to address these shortcomings is underway and will be presented elsewhere.

Performing the phase canonicalization ensures that using identity for $\{\mathbf{U}_{\mathbf{k}}\}$ produces well-localized WFs for many bands. Note, however, that the ``intra-cell'' localization is not assisted by the phase canonicalization; thus, it alone will not be sufficient for systems with large unit cells. To be able to locate the global maxima of the PMWF functional by sampling the initial guesses, by default we initialize $\{\mathbf{U}_{\mathbf{k}}\}$ with a (quasi)random unitary matrix generated from a user-supplied seed. Note that the same unitary matrix is used for every $\mathbf{k}$ in order to preserve the benefit of phase canonicalization.

All computations reported in the manuscript used the CPR guess generated with the same (default) seed value. In the Supporting Information, we report additional computations that, after the phase canonicalization, initialized $\{\mathbf{U}_{\mathbf{k}}\}$ with identity matrices as well as with random $\{\mathbf{U}_{\mathbf{k}}\}$ generated nonuniformly across the first Brillouin zone (i.e., a different quasirandom matrix was used for every $\mathbf{k}$ quadrature point, thereby canceling the benefit of phase canonicalization). The performance of the default CPR and ``identity'' CPR was found to be similar, whereas using the random nonuniform guess required significantly more iterations to reach convergence; however, the final PM functional value was found to be the same for all initial guesses.

Clearly, since the CPR method is defined intrinsically without any reference to the LCAO representation, it can be utilized in the context of non-LCAO (PW and other) representations as long as the Bloch orbitals can be projected onto the MBS AOs. As discussed above, projection on localized states is common in preparing trial Wannier functions;\cite{VRG:marzari:1997:PRB,VRG:zicovich-wilson:2001:JCP,VRG:mustafa:2015:PRB} the use of a full {\em minimal} AO basis for projection makes CPR guess (1) well defined even in the limit of a complete orbital basis set (in contrast to the approach of Ref. \citenum{VRG:zicovich-wilson:2001:JCP}) and (2) more black-box by eliminating the need to guess positions or composition of target Wannier functions (in contrast to the approaches of Refs. \citenum{VRG:marzari:1997:PRB,VRG:mustafa:2015:PRB}).

\subsubsection{PM Functional Gradient}
To find a maximum of the PM functional, we will use global gradient-based optimization. The gradient of $P$ with respect to $\{\mathbf{U}_{\mathbf{k}}\}$
is expressed straightforwardly:
\begin{align}
\label{grad}
\frac{\partial P}{\partial (U^{i}_{j,{\bf k}})^{*}}
= &
\frac{p}{\sqrt{N_{k}}} \sum_{A,\mathbf{R}} |\overline{Q}_{i}^{A_\mathbf{R}}|^{p-1}
\sum_{\mu \in A} \left[
\bra{\co{j}{\mathbf{k}}}\ket{\bar{\phi}_{\mu,\mathbf{R}}} \overline{W}^i_{\mu,\mathbf{R}} + \left(\overline{C}^{j,\mathbf{k}}_{\mu,\mathbf{R}}\right)^* \bra{\bar{\phi}_{\mu,\mathbf{R}}}\ket{\wann{i}{\mathbf{0}}} 
\right]
\nonumber \\
= & \frac{p}{\sqrt{N_{k}}} \sum_{A,\mathbf{R}} |\overline{Q}_{i}^{A_\mathbf{R}}|^{p-1}
e^{-\mathrm{i}\mathbf{k}\cdot\mathbf{R}} \sum_{\mu \in A} \left[
 \bra{\co{j}{\mathbf{k}}}\ket{\bar{\phi}_{\mu,\mathbf{0}}} \overline{W}^i_{\mu,\mathbf{R}} +  \left(\overline{C}^{j,\mathbf{k}}_{\mu,\mathbf{0}}\right)^* \bra{\bar{\phi}_{\mu,\mathbf{R}}}\ket{\wann{i}{\mathbf{0}}} 
\right]
\end{align}
\cref{grad} uses the standard complex-valued form\cite{Brandwood1983} of the derivative of a real-valued function of complex-valued parameters, $(\partial f / \partial z^*) \equiv (\partial f/ \partial \Re z) + \mathrm{i} \, (\partial f/ \partial \Im z)$, which makes the notation more compact. The evaluation of the gradient again leverages the Bloch-MBS overlaps (\cref{eq:Smbsco}) and is straightforward.

It is of course more convenient to express the PMWF functional in terms of nonredundant variables by introducing the standard exponential parametrization of a unitary matrix, $\mathbf{U} \equiv \exp(\bm{\kappa} - \bm{\kappa}^\dagger)$, where $\bm{\kappa}$ is a complex triangular matrix. It is straightforward to convert the ``Euclidean'' gradient arranged as a matrix for each $\mathbf{k}$,
\begin{equation}
\left( \mathbf{\Gamma}_{\mathbf{k}} \right)^i_j
\equiv 
\frac{\partial P}{\partial (U^i_{j,{\bf k}})^{*}}
\end{equation}
to its ``curvilinear'' counterpart,
\begin{align}
\left( \mathbf{G}_{\mathbf{k}} \right)^i_j \equiv \frac{\partial P}{\partial (\kappa^i_{j,{\bf k}})^{*}}
\end{align}
as follows:\cite{VRG:abrudan:2008:ITSP,abrudan}
\begin{equation}
\mathbf{G}_{\mathbf{k}} =
(\mathbf{U}_{\mathbf{k}})^{\dagger}\mathbf{\Gamma}_{\mathbf{k}}
-(\mathbf{\Gamma}_{\mathbf{k}})^{\dagger} \mathbf{U}_{\mathbf{k}}.
\end{equation}
Note that $\mathbf{G}_{\mathbf{k}}$ is antihermitian, just like $\bm{\kappa}_{\bf k} - \bm{\kappa}_{\bf k}^\dagger$.
Maximization of the PMWF functional expressed in terms of $\bm{\kappa}_{\bf k}$ is a standard (unconstrained) nonlinear optimization problem; its solution is described in the following section.

\subsubsection{Direction Choice: SA, CG, BFGS}
The main focus on our work is how to solve for the PMWFs robustly. As the reference methods for locating the PM functional maxima, we will use  the steepest ascent (SA)
and (nonlinear) conjugate gradient (CG) methods. In particular, we have chosen
three specific varieties of CG for comparison: the
Polak-Ribi\`ere formulation (CGPR),\cite{cgpr} the Fletcher-Reeves formulation (CGFR),\cite{cgfr}
and the Hestenes-Stiefel formulation (CGHS).\cite{cghs} Due to the well-known
nature of SA and nonlinear CG (see, for example, any textbook on numerical optimization), 
we will not discuss their implementation details here,
except to note that, for each of the three different
CG variants considered, we also varied the number
of SA steps taken before beginning CG.
The numbers of initial SA steps considered in this work were
1, 2, 5, 10, and 15, meaning that, for each system,
we ran a total of 15 different CG calculations.
Also note that the optimization problem of the real-valued PMWF functional $P$ was recast (as usual) in terms of real and imaginary components of the complex-valued parameters $U^i_{j,\mathbf{k}}$, i.e., henceforth the gradient and other vectors will consist of $o (o-1) N_k$ real numbers only, where $o$ is the number of Bloch orbitals being localized;\footnote{In practice the implementation uses pairs of real square matrices to represent $\bm{\kappa}_{\bf k}$, $\mathbf{G}_{\mathbf{k}}$, thus using $2 o^2 N_k$ (instead of $o (o-1) N_k$) parameters. This is due to the lack of support for the (anti)symmetric matrix format in the Eigen library used for numerical manipulations in the PM solver.} complex-valued formulations of the optimization problem\cite{VRG:sorber:2012:SJO} were not considered here.

The BFGS method,
though also well-known, warrants a bit of discussion. In particular,
we have employed the ``two-loop recursion'' form of the limited-memory BFGS\cite{VRG:nocedal:1980:MC} (L-BFGS; henceforth we will omit the ``L-'' prefix unless this algorithmic detail is relevant) algorithm for updating
the estimated inverse Hessian; the initial estimate of the inverse Hessian was chosen to be an identity matrix. Because each BFGS iteration depends on some number of prior iterations (the ``history'') to generate
an updated estimate of the inverse Hessian,
it is necessary to select the size of this history (i.e., the number of iterations kept).
In addition, regardless of the history size, the first update must be necessarily
be steepest ascent since the history does not yet exist. Of course, it is also possible
to perform any number of SA steps before beginning the BFGS procedure, and it
is these two parameters (the history size and the number of initial SA steps)
that define the BFGS algorithm as implemented here.
For all systems, we used five different initial SA values and five different
history sizes, for a total of 25 BFGS solver setups.
The initial SA values considered were 1, 2, 5, 10, and 15; the history sizes considered were also 1, 2, 5, 10, and 15. In future discussion, BFGS parameters are indicated as
$\langle \text{No. of SA steps}, \text{history size}\rangle$.

\subsubsection{Line Search}
Regardless of how the direction was chosen (SA, CG, BFGS), the 
line search was performed in the same manner, using a low-order polynomial approximation of the objective function along the trial direction. First, the proposed direction is checked to point uphill (if not, the trial direction is reversed). Then, given a fitting range upper bound $T_\mu$ (see below) and
the polynomial order $n=4$, the PM functional is evaluated at $n+1$ evenly-spaced points, $\{\mu_0\equiv0, \mu_1\equiv T_\mu/n, \mu_2\equiv 2T_\mu/n \dots \}$, in [0, $T_\mu$]. The $\{P(\mu_{i})\}$ set is then used to construct a polynomial fit, $f(\mu_{i})$, and the bisection method is used to find a root of $f'(\mu_{i})$.
The fitting range and polynomial orders are determined as follows.
\begin{itemize}
    \item Iteration 0: $T_{\mu}$ is estimated from the shortest orbital rotation frequency $\omega_\mathrm{max}$ along the given direction via Eq. (15) of Ref. \citenum{abrudan} (the largest $\omega_\mathrm{max}$ is chosen among all \textbf{k} points).
    \item Iteration $i$: The upper bound from the previous iteration is used as a trial upper bound. If $P(\mu_1)$ is less than $P(0)$ then $T_\mu$ is reset to $\mu_1$, else $P(\mu_k)$ are evaluated for increasing $k$ until $P(\mu_k) < P(\mu_{k-1})$ is found. If such $k$ is found, values $\{P(\mu_{i})\}, i\in[0,k]$ are fitted to polynomial of order $k-1$; else $T_\mu$ is reset to $5T_\mu$ and the process is repeated.
\end{itemize}
Finally, if we fail to find an acceptable upper bound along a chosen direction, we will reset to the SA direction. If the upper bound is not located in the SA direction or the root finder fails, then the upper bound is recomputed as done at the start. These resets are rarely necessary when performing BFGS.

Note that, in addition to resetting the CG direction to the SA direction whenever an
acceptable upper bound cannot be found, it is also necessary to reset the
CG direction every $n$ iterations, where $n$ is the number of orbitals being
localized. This is because a system with $n$ variables can only have $n$
conjugate directions.

\section{Computational Details}
All calculations were carried out in the developmental version of the Massively Parallel Quantum Chemistry (MPQC) package (version 4.0.0).\cite{mpqc} 
All computations
were performed on the NewRiver commodity cluster at Virginia Tech.\cite{VTARCNewRiver}

Hartree-Fock computations were carried out using the reduced-scaling
LCAO formalism described in Ref. \citenum{VRG:wang:2020:JCP}. The Coulomb potential was evaluated using multipole-accelerated real-space summation and density-fitting approximation, whereas the exchange potential was evaluated using concentric atomic density fitting.\cite{VRG:wang:2020:JCP}
Table~\ref{crysData} lists the test systems as well as the corresponding orbital basis set,
the Monkhorst-Pack mesh size, and the PM convergence threshold employed for each. In all cases, the def2-SVP-J basis set\cite{weigend2006} was used as the density
fitting basis. The pseudoinverse atomic charges were evaluated using
the Huzinaga MINI basis set\cite{MINI:H,MINI,BSE} as the minimal AO basis. All occupied orbitals (including core) were localized. The PM functional with $p=4$ (see \cref{P}) was utilized throughout. Complete input files (which specify unit cell parameters) and
geometries can be found in the Supporting Information.

Note that the convergence thresholds that we use in this work are significantly tighter than the typical threshold of $10^{-5}$ utilized in the comparable studies.\cite{VRG:hoyvik:2012:JCTC,VRG:lehtola:2013:JCTC} A slightly looser convergence in bulk Si was due to the greater range of the lengthscales of the Wannier functions in that system; localizing core and valence orbitals separately would alleviate these issues, but was not pursued in order to keep the solver assessment protocol as uniform and as stringent as possible. 

\begin{table}[!htb]
\begin{center}
\begin{tabular}{llccc}
\hline\hline
System & OBS & Monkhorst-Pack mesh size$^a$ & PM gradient norm \\
\hline
trans-$(\text{C}_{2}\text{H}_{2})_\infty$ & 6-31G* & 101 & $10^{-8}$ \\
$(\text{C}_{2}\text{H}_{4})_\infty$ & 6-31G* & 101 & $10^{-8}$ \\
(4,0) nanotube & 6-31G* & 51 & $10^{-8}$ \\
Graphene & 6-31G & $21 \times 21$ & $10^{-8}$ \\
h-BN & 6-31G & $21 \times 21$ & $10^{-8}$ \\
LiH & CR-cc-pVDZ\cite{VRG:lorenz:2012:JCP} & $11 \times 11 \times 11$ & $10^{-8}$ \\
Diamond & 6-31G* & $11 \times 11 \times 11$ & $10^{-8}$ \\
Silicon & 6-31G* & $11 \times 11 \times 11$ & $10^{-7}$ \\
\hline\hline
\end{tabular}
\caption{Test systems used to assess the performance of PMWF solvers, along with the relevant computational details.}
\label{crysData}
$^a$ All systems except h-BN utilized primitive unit cells; an orthorhombic non-primitive unit cell was used for h-BN.
\end{center}
\end{table}

\section{Results}
As discussed in \cref{sec:initialguess}, the PM functional does not specify WFs uniquely, thus to compare computed WFs we compare the corresponding values of the PM functional; two WF sets will be referred to as {\em equivalent} if they correspond to the same value of the PM functional up to target precision. For every test system, all PMWF solvers considered in this work (BFGS with 25 parameter settings,
CG with 15 parameter settings, and SA; see \cref{sec:solvers}), when converged
successfully, produced WFs that were practically equivalent;
only for silicon and the carbon nanotube did the variance of the final $P$ value
exceeded $10^{-8}$
(see the
Supporting Information for more details). Limited testing also indicated that the use
of the standard (CPR) and nonstandard guesses produced equivalent sets of Wannier
functions.


\begin{table}[!htb]
\scriptsize
\begin{center}
\begin{tabular}{cccccc|cccccc}
\hline\hline
System & Solver & Min & Max & Mean & St. Dev. & System & Solver & Min & Max & Mean & St. Dev. \\
\hline
\multirow{5}{*}{$\text{C}_{2}\text{H}_{2}$} & BFGS & 31 & 47 & 40.5 & 3.5 & \multirow{5}{*}{h-BN} & BFGS & 45 & 57 & 50.2 & 3.3 \\
& CG & 1398 & 2534 & 2007.5 & 389.1 && CG & 174 & 808 & 260.8 & 160.9 \\
& CGPR & 1398 & 1934 & 1708.6 & 209.0 && CGPR$^c$ & 174 & 808 & 339.3 & 312.7 \\
& CGFR & 1752 & 1832 & 1801.0 & 35.5 && CGFR & 174 & 250 & 210.6 & 35.5 \\
& CGHS & 2483 & 2534 & 2513.0 & 22.5 && CGHS & 235 & 257 & 248.2 & 8.3 \\
& SA & 1745 & -- & -- & -- && SA & 175 & -- & -- & -- \\
\hline
\multirow{5}{*}{$\text{C}_{2}\text{H}_{4}$} & BFGS & 24 & 37 & 29.0 & 3.2 & \multirow{5}{*}{LiH} & BFGS & 10 & 22 & 14.5 & 4.2 \\
& CG & 66 & 729 & 230.7 & 209.8 && CG & 14 & 1441 & 783.2 & 607.2 \\
& CGPR & 69 & 425 & 215.0 & 133.2 && CGPR & 786 & 1062 & 900.6 & 122.7 \\
& CGFR & 109 & 187 & 158.6 & 33.8 && CGFR & 14 & 23 & 17.6 & 3.8 \\
& CGHS & 66 & 729 & 318.4 & 344.6 && CGHS & 1395 & 1441 & 1431.4 & 20.4 \\
& SA & 52 & -- & -- & -- && SA & 1457 & -- & -- & -- \\
\hline
\multirow{5}{*}{Diamond} & BFGS & 23 & 28 & 24.7 & 1.4 & \multirow{5}{*}{Nanotube} & BFGS & 75 & 101 & 86.1 & 8.3 \\
& CG & 39 & 212 & 89.6 & 65.3 && CG & 713 & 2275 & 1174.5 & 471.0 \\
& CGPR & 43 & 212 & 133.0 & 69.6 && CGPR & 935 & 2275 & 1309.0 & 581.3 \\
& CGFR & 39 & 51 & 46.2 & 5.1 && CGFR & 1015 & 1903 & 1447.8 & 334.5 \\
& CGHS$^a$ & -- & -- & -- & -- && CGHS & 713 & 853 & 766.8 & 55.4 \\
& SA$^b$ & -- & -- & -- & -- && SA & 485 & -- & -- & -- \\
\hline
\multirow{5}{*}{Graphene} & BFGS & 53 & 186 & 86.0 & 42.0 & \multirow{5}{*}{Silicon} & BFGS & 21 & 29 & 24.1 & 2.0 \\
& CG & 2657 & 3666 & 3296.8 & 275.8 && CG & 32 & 2975 & 954.9 & 1240.7 \\
& CGPR$^c$ & 2657 & 3666 & 3234.8 & 439.4 && CGPR & 113 & 221 & 176.4 & 44.3 \\
& CGFR & 3321 & 3376 & 3346.4 & 21.6 && CGFR & 32 & 83 & 55.4 & 19.9 \\
& CGHS$^a$ & -- & -- & -- & -- && CGHS & 2243 & 2975 & 2633.0 & 310.8 \\
& SA$^b$ & -- & -- & -- & -- && SA & 3171 & -- & -- & -- \\
\hline\hline
\end{tabular}
\caption{Length Summary}
\label{lengthSumm}
\end{center}
$^a$ All five calculations failed to converge. \\
$^b$ Calculation failed to converge. \\
$^c$ One calculation failed to converge.
\end{table}

Although all solvers produced equivalent sets of PMWFs, the number of iterations needed
to locate the maxima of the PM functional differed dramatically between the different
classes of solvers. Column ``Min'' in \cref{lengthSumm} lists the minimum number of
iterations needed to arrive at the solution, broken down by the system and solver
class. In all cases except the zero-gap system, the BFGS solver converged in fewer than
60 iterations in the best-case scenario, and only two systems required
more than 100 iterations in the worst-case scenario.
Each of the three CG variants managed to converge in under 100 iterations for at
least one system, but even CGFR, which saw the most success in converging
in less than 100 iterations, only did so for three systems. In other cases,
the various CG variants could take thousands of iterations to converge, and
CGHS failed to converge in 4,000 iterations for both diamond and graphene,
regardless of the starting number of steepest ascent steps.
Depending on the system, SA could converge in under 50 iterations or take
hundreds or thousands to converge; in two cases, it failed to converge
at all within 4,000 iterations. Even in the best-case scenario (i.e., minimum
iterations to convergence), BFGS was always superior to CG and SA, though
the latter two could sometimes come close. But even when CG and SA
were nearly comparable to BFGS in terms of number of iterations to solution,
their convergence behavior could not be counted on, as is evidenced
by the much larger calculation length standard deviations for these
solvers compared with the same metric for BFGS. Overall, the performance
of the SA and CG solvers can, at best, be characterized as unreliable;
in contrast, BFGS is reliably rapid.


\begin{figure}
\subfloat[trans-$(\text{C}_{2}\text{H}_{2})_\infty$]{%
  \includegraphics[width=0.7\textwidth]{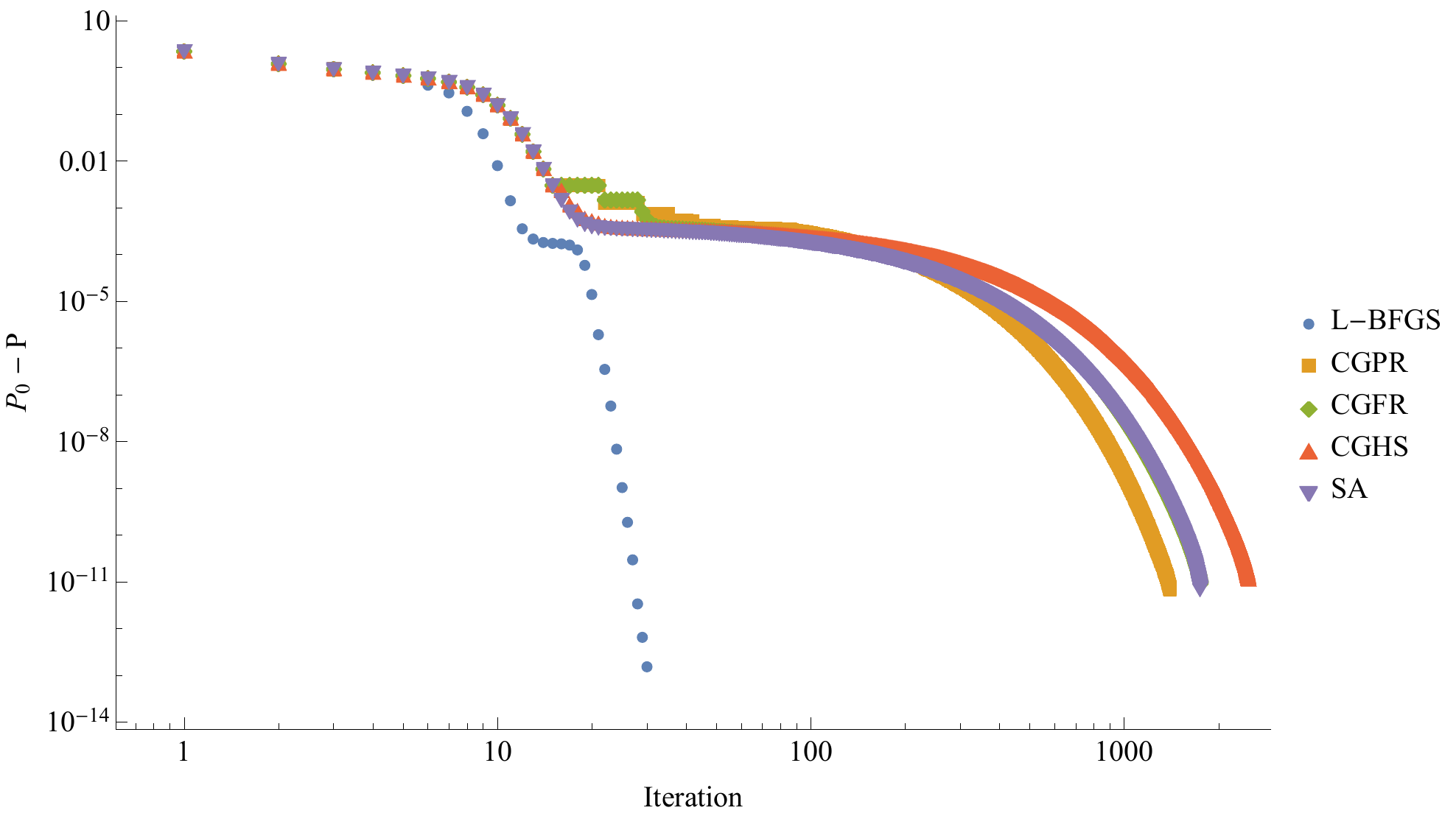}
}

\subfloat[h-BN]{%
  \includegraphics[width=0.7\textwidth]{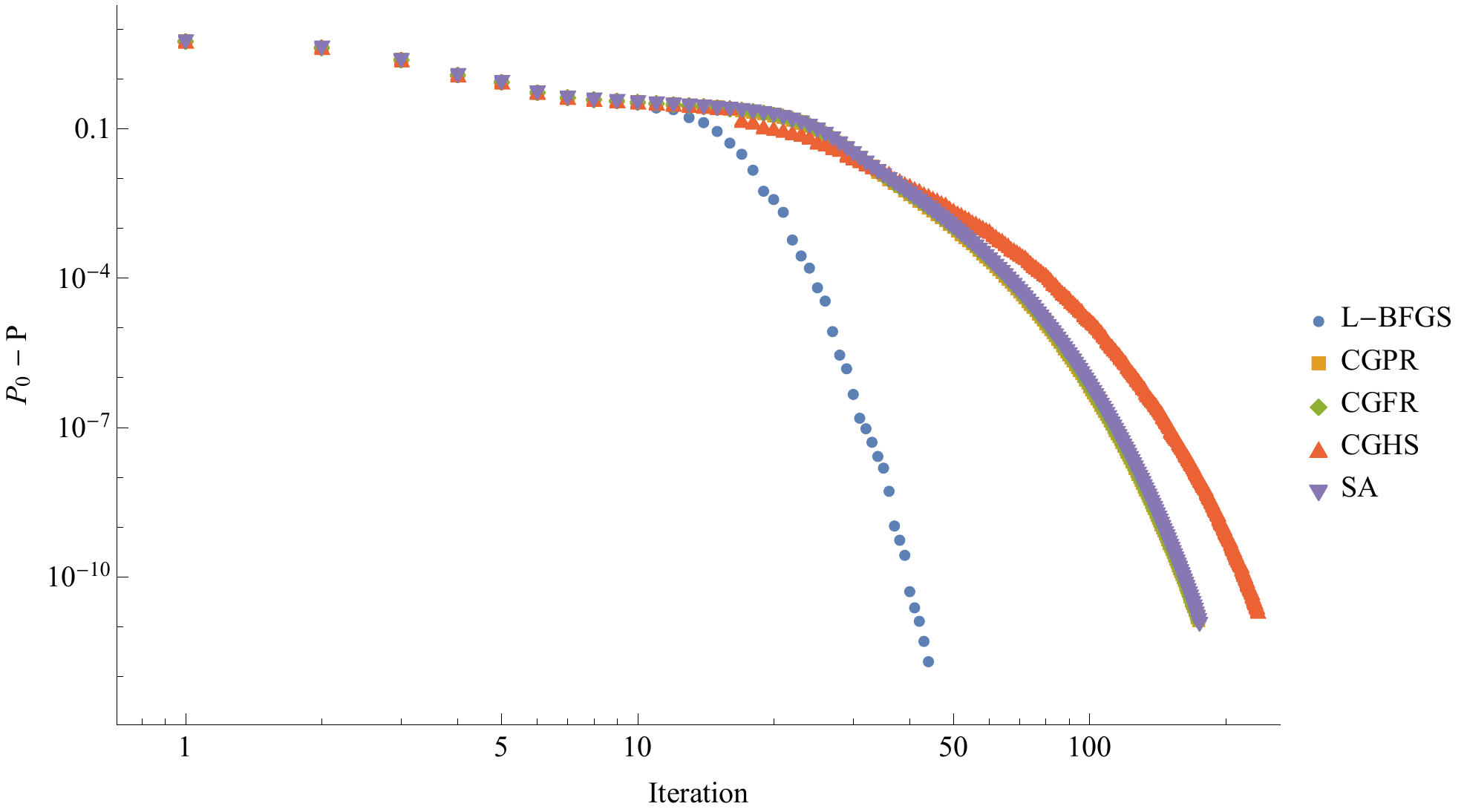}
}

\subfloat[LiH]{%
  \includegraphics[width=0.7\textwidth]{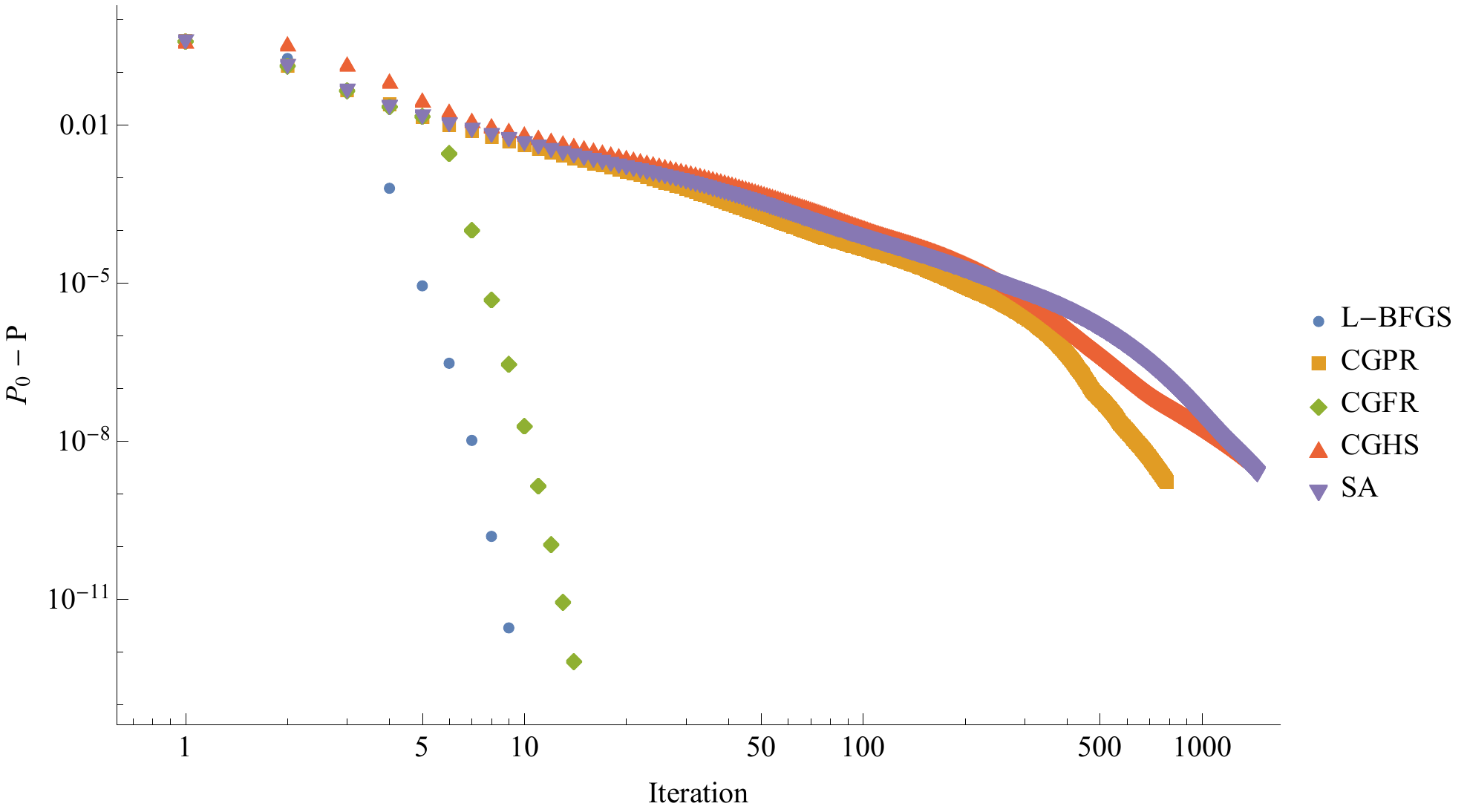}
}
\caption{Plots of the difference between the PM functional value $P$ and
the converged value $P_{0}$ vs the iteration count. For each system represented, the shortest calculation in each category
was chosen; $P_{0}$ was taken to be the greatest final $P$ value across the selected calculations for a given
system. \label{fig:conv}}
\end{figure}

Due to the significant variation in the performance of the CG and SA solvers for different systems, it is difficult to pinpoint the origin of their struggles. \cref{fig:conv} illustrates the convergence patterns observed for representative 1-d, 2-d, and 3-d test systems. The superior performance of BFGS compared to CG and SA is plainly visible. Also note the extended plateau exhibited by the CG and SA solvers in the 1-d system, which is a typical CG convergence pattern\cite{VRG:axelsson:2003:MaCiSa} and which suggests a high condition number of the PM Hessian in this system. The BFGS solver also exhibits a plateau in this system, but it is much shorter. Lastly, the quality of the initial guess can vary greatly from system to system: in the 1-d system, the initial guess is clearly significantly worse than in the 2-d and 3-d systems, as indicated by significantly larger initial deviations.

The performance of BFGS is also relatively insensitive to the choice of its parameters, namely the number of SA steps at the start and the history size, as illustrated in 
\cref{lengthSumm}. The small standard deviation of the BFGS solvers' performance ($<10$ for all systems other than graphene) indicates that approximately the same number of iterations is needed to locate the maximum regardless of the BFGS parameter values.
In contrast, the CG solvers' performance can depend strongly on the number of starting SA steps. This correlates with the unreliable performance of the CG solvers that we noted previously.

In addition, the number of iterations needed for the CG solver also correlates
as expected with the findings of J\'onsson et al.,\cite{VRG:jonsson:2017:JCTC}
whose implementation of CG required hundreds of iterations in periodic systems.
Cases where our CG implementation converged in fewer iterations may be a
result of differences in initial guess (CPR in our work, random guess in their work),
while situations in which our CG implementation took longer may be
due to the tighter convergence thresholds employed herein.
Furthermore, the only
3-d system that J\'onsson et al. considered was a benzene
crystal,\cite{VRG:jonsson:2017:JCTC} which, like other molecular crystals, would have
low-dispersion band structure and for which it should be easier to generate a localized
initial guess (see our discussion in \cref{sec:initialguess}). We focused on the more
challenging ionic and covalent 3-d systems.


\begin{table}[!htb]
\begin{center}
\begin{tabular}{ccccc}
\hline\hline
$\langle$ \# of SA, History $\rangle$ & Min & Max & Mean & St. Dev \\
\hline
$\langle$1, 1$\rangle$ & 1.00000 & 2.83019 & 1.33181 & 0.61693 \\
$\langle$1, 2$\rangle$ & 1.00000 & 1.81132 & 1.19016 & 0.27206 \\
$\langle$1, 5$\rangle$ & 1.04348 & 1.35849 & 1.20877 & 0.11738 \\
$\langle$1, 10$\rangle$ & 1.03774 & 1.41935 & 1.19625 & 0.14074 \\
$\langle$1, 15$\rangle$ & 1.01887 & 1.38095 & 1.18609 & 0.12330 \\
$\langle$2, 1$\rangle$ & 1.00000 & \redBox{3.50943} & 1.41816 & \redBox{0.85422} \\
$\langle$2, 2$\rangle$ & 1.00000 & 1.43396 & 1.14882 & 0.13846 \\
$\langle$2, 5$\rangle$ & 1.04348 & \blueBox{1.22581} & \blueBox{1.13897} & \blueBox{0.06151} \\
$\langle$2, 10$\rangle$ & 1.03774 & 1.41935 & 1.14482 & 0.12524 \\
$\langle$2, 15$\rangle$ & 1.01887 & \greenBox{1.19355} & \greenBox{1.11134} & \greenBox{0.05189} \\
$\langle$5, 1$\rangle$ & 1.00000 & \orangeBox{3.39623} & 1.43594 & \orangeBox{0.80097} \\
$\langle$5, 2$\rangle$ & 1.00000 & 2.05660 & 1.22799 & 0.34514 \\
$\langle$5, 5$\rangle$ & 1.04000 & 1.41509 & 1.17032 & 0.13938 \\
$\langle$5, 10$\rangle$ & 1.05660 & 1.41935 & 1.17322 & 0.12475 \\
$\langle$5, 15$\rangle$ & 1.01887 & 1.30000 & 1.15381 & 0.09378 \\
$\langle$10, 1$\rangle$ & 1.00000 & 1.70000 & 1.26942 & 0.28516 \\
$\langle$10, 2$\rangle$ & 1.00000 & 1.88679 & 1.28996 & 0.33131 \\
$\langle$10, 5$\rangle$ & 1.00000 & 1.70000 & 1.21055 & 0.25964 \\
$\langle$10, 10$\rangle$ & 1.01887 & 1.70000 & 1.23150 & 0.23658 \\
$\langle$10, 15$\rangle$ & 1.03774 & 1.80000 & 1.24892 & 0.24966 \\
$\langle$15, 1$\rangle$ & 1.04444 & 3.01887 & \redBox{1.60319} & 0.66128 \\
$\langle$15, 2$\rangle$ & \redBox{1.10667} & 2.30189 & \orangeBox{1.45403} & 0.47729 \\
$\langle$15, 5$\rangle$ & \orangeBox{1.06667} & 2.10000 & 1.32343 & 0.33587 \\
$\langle$15, 10$\rangle$ & 1.05660 & 2.10000 & 1.29656 & 0.34030 \\
$\langle$15, 15$\rangle$ & 1.05660 & 2.20000 & 1.33176 & 0.36673 \\
\hline\hline
\end{tabular}
\caption{Statistical analyses of the number of iterations to solution for each BFGS solver parameter set. Colored boxes identify the smallest and largest values in each column as follows:
green (smallest), blue (second smallest), red (largest), orange (second largest).}
\label{ratioAnal}
\end{center}

$^a$ A ``Min'' value of 1 indicates that this solver was the fastest (required the fewest iterations) for at least 1 test system. \\
$^b$ A ``Max'' value of 1.5 means that, for every test system, this solver required at most 50\% more iterations than the fastest solver for that system.\\
$^c$ A ``Mean'' value of 1.2 means that, for a give test system, this solver required on average 20\% more iterations than the fastest solver for that system.
\end{table}

Although the BFGS-based PMWF solver is already highly robust, some parameter choices are systematically better than others.
Thus, we analyzed the distribution of the number of iterations needed to locate the solution for a given system with the given BFGS solver parameter values relative to the {\em smallest} number of iterations needed for that system; the results of this analysis are listed in \cref{ratioAnal} (see the Supporting Information for the raw number of iterations for each system). We highlighted the smallest and largest values in each
column to make it easier to locate the fastest and slowest solvers. The $\langle2, 15\rangle$ BFGS solver is overall the fastest, both on average and in the worst-case
scenario, and thus is the recommended default choice. 


\section{Summary and Perspective}
We described a robust BFGS-based solver to obtain (generalized) Pipek-Mezey Wannier functions in periodic solids whose use was pioneered recently by J\'onsson et al.\cite{VRG:jonsson:2017:JCTC} The PM functional used in this work utilized atomic charges using a simple pseudoinverse projection onto a pre-defined minimal AO basis, thus making its evaluation convenient in a periodic LCAO representation. An essential contributor to the robustness of the solver is the novel automated ``Canonicalize Phase then Randomize'' (CPR) method for generating the initial guess. The limited-memory BFGS solver converged very tightly in fewer than 60 iterations in 1-, 2-, and 3-dimensional systems featuring a variety of bonding patterns (covalent, ionic) and gaps, even in systems with very large BvK unit cells (thousands of atoms). The sole exception was one system with a vanishing gap where $\sim80$ iterations were needed. This is a significant improvement on the more traditional SA-based solver that can require hundreds or thousands of iterations, or the nonlinear CG solvers that often converge faster than SA, but can unpredictably converge very slowly or even fail to converge at all. Although the performance of the solver was relatively insensitive to the BFGS history size, the near-optimal choice of history size was determined to be 1, making the BFGS solver notionally similar in the operation and storage costs to that of CG.

Clearly, the BFGS-based solver should be robustly usable for computing other generalized Wannier functions, such as the Boys (maximally-localized) WFs. Although here we only explored localization of occupied states, the solver should be also applicable to the unoccupied states with valence character. The automated CPR method for generating initial WF guesses could be used in conjunction with PW-based representations of Bloch orbitals, potentially improving on the existing approaches\cite{VRG:marzari:1997:PRB,VRG:mustafa:2015:PRB}. Lastly, it is also worthwhile to assess the efficacy of BFGS-based solvers for other challenging orbital optimization problems in molecules and solids, such as localization of unoccupied (virtual) orbitals\cite{VRG:hoyvik:2012:JCTC,VRG:lehtola:2013:JCTC} and for the orbital optimization in the context of Perdew-Zunger self-interaction-corrected DFT (notably, some limited use of BFGS in this context was recently reported by Lehtola et al.\cite{VRG:lehtola:2016:JCTC}).

\begin{acknowledgement}
This work was supported by the U.S. National Science Foundation (awards 1550456 and 1800348). We also acknowledge Advanced Research Computing at Virginia Tech (www.arc.vt.edu) for providing computational resources and technical support that have contributed to the results reported within this paper. The Flatiron Institute is a division of the Simons Foundation.
\end{acknowledgement}

~\\
{\bf {\Large Supporting Information}}\\
The complete computation input files and geometries,
select  raw computational data analyzed in the manuscript.

\bibliography{references,vrgrefs}

\providecommand{\latin}[1]{#1}
\makeatletter
\providecommand{\doi}
  {\begingroup\let\do\@makeother\dospecials
  \catcode`\{=1 \catcode`\}=2 \doi@aux}
\providecommand{\doi@aux}[1]{\endgroup\texttt{#1}}
\makeatother
\providecommand*\mcitethebibliography{\thebibliography}
\csname @ifundefined\endcsname{endmcitethebibliography}
  {\let\endmcitethebibliography\endthebibliography}{}
\begin{mcitethebibliography}{56}
\providecommand*\natexlab[1]{#1}
\providecommand*\mciteSetBstSublistMode[1]{}
\providecommand*\mciteSetBstMaxWidthForm[2]{}
\providecommand*\mciteBstWouldAddEndPuncttrue
  {\def\EndOfBibitem{\unskip.}}
\providecommand*\mciteBstWouldAddEndPunctfalse
  {\let\EndOfBibitem\relax}
\providecommand*\mciteSetBstMidEndSepPunct[3]{}
\providecommand*\mciteSetBstSublistLabelBeginEnd[3]{}
\providecommand*\EndOfBibitem{}
\mciteSetBstSublistMode{f}
\mciteSetBstMaxWidthForm{subitem}{(\alph{mcitesubitemcount})}
\mciteSetBstSublistLabelBeginEnd
  {\mcitemaxwidthsubitemform\space}
  {\relax}
  {\relax}

\bibitem[Boys(1960)]{VRG:boys:1960:RMP}
Boys,~S.~F. Construction of {{Some Molecular Orbitals}} to {{Be Approximately
  Invariant}} for {{Changes}} from {{One Molecule}} to {{Another}}. \emph{Rev.
  Mod. Phys.} \textbf{1960}, \emph{32}, 296--299\relax
\mciteBstWouldAddEndPuncttrue
\mciteSetBstMidEndSepPunct{\mcitedefaultmidpunct}
{\mcitedefaultendpunct}{\mcitedefaultseppunct}\relax
\EndOfBibitem
\bibitem[Foster and Boys(1960)Foster, and Boys]{FB}
Foster,~J.~M.; Boys,~S.~F. Canonical Configurational Interaction Procedure.
  \emph{Rev. Mod. Phys.} \textbf{1960}, \emph{32}, 300--302\relax
\mciteBstWouldAddEndPuncttrue
\mciteSetBstMidEndSepPunct{\mcitedefaultmidpunct}
{\mcitedefaultendpunct}{\mcitedefaultseppunct}\relax
\EndOfBibitem
\bibitem[Pipek and Mezey(1989)Pipek, and Mezey]{VRG:pipek:1989:JCP}
Pipek,~J.; Mezey,~P.~G. A Fast Intrinsic Localization Procedure Applicable for
  {\emph{Ab}} {\emph{Initio}} and Semiempirical Linear Combination of Atomic
  Orbital Wave Functions. \emph{J. Chem. Phys.} \textbf{1989}, \emph{90},
  4916--4926\relax
\mciteBstWouldAddEndPuncttrue
\mciteSetBstMidEndSepPunct{\mcitedefaultmidpunct}
{\mcitedefaultendpunct}{\mcitedefaultseppunct}\relax
\EndOfBibitem
\bibitem[Edmiston and Ruedenberg(1963)Edmiston, and Ruedenberg]{ER}
Edmiston,~C.; Ruedenberg,~K. Localized Atomic and Molecular Orbitals.
  \emph{Rev. Mod. Phys.} \textbf{1963}, \emph{35}, 457--464\relax
\mciteBstWouldAddEndPuncttrue
\mciteSetBstMidEndSepPunct{\mcitedefaultmidpunct}
{\mcitedefaultendpunct}{\mcitedefaultseppunct}\relax
\EndOfBibitem
\bibitem[{von Niessen}(1972)]{VRG:vonniessen:1972:JCP}
{von Niessen},~W. Density {{Localization}} of {{Atomic}} and {{Molecular
  Orbitals}}. {{I}}. \emph{J. Chem. Phys.} \textbf{1972}, \emph{56},
  4290--4297\relax
\mciteBstWouldAddEndPuncttrue
\mciteSetBstMidEndSepPunct{\mcitedefaultmidpunct}
{\mcitedefaultendpunct}{\mcitedefaultseppunct}\relax
\EndOfBibitem
\bibitem[Cioslowski(1991)]{VRG:cioslowski:1991:JMC}
Cioslowski,~J. Partitioning of the Orbital Overlap Matrix and the Localization
  Criteria. \emph{J Math Chem} \textbf{1991}, \emph{8}, 169--178\relax
\mciteBstWouldAddEndPuncttrue
\mciteSetBstMidEndSepPunct{\mcitedefaultmidpunct}
{\mcitedefaultendpunct}{\mcitedefaultseppunct}\relax
\EndOfBibitem
\bibitem[Alcoba \latin{et~al.}(2006)Alcoba, Lain, Torre, and
  Bochicchio]{VRG:alcoba:2006:JCC}
Alcoba,~D.~R.; Lain,~L.; Torre,~A.; Bochicchio,~R.~C. An Orbital Localization
  Criterion Based on the Theory of ``Fuzzy'' Atoms. \emph{J. Comput. Chem.}
  \textbf{2006}, \emph{27}, 596--608\relax
\mciteBstWouldAddEndPuncttrue
\mciteSetBstMidEndSepPunct{\mcitedefaultmidpunct}
{\mcitedefaultendpunct}{\mcitedefaultseppunct}\relax
\EndOfBibitem
\bibitem[Knizia(2013)]{VRG:knizia:2013:JCTC}
Knizia,~G. Intrinsic {{Atomic Orbitals}}: {{An Unbiased Bridge}} between
  {{Quantum Theory}} and {{Chemical Concepts}}. \emph{J. Chem. Theory Comput.}
  \textbf{2013}, \emph{9}, 4834--4843\relax
\mciteBstWouldAddEndPuncttrue
\mciteSetBstMidEndSepPunct{\mcitedefaultmidpunct}
{\mcitedefaultendpunct}{\mcitedefaultseppunct}\relax
\EndOfBibitem
\bibitem[Lehtola and J{\'o}nsson(2014)Lehtola, and
  J{\'o}nsson]{VRG:lehtola:2014:JCTC}
Lehtola,~S.; J{\'o}nsson,~H. Pipek\textendash{{Mezey Orbital Localization Using
  Various Partial Charge Estimates}}. \emph{J. Chem. Theory Comput.}
  \textbf{2014}, \emph{10}, 642--649\relax
\mciteBstWouldAddEndPuncttrue
\mciteSetBstMidEndSepPunct{\mcitedefaultmidpunct}
{\mcitedefaultendpunct}{\mcitedefaultseppunct}\relax
\EndOfBibitem
\bibitem[Janowski(2014)]{VRG:janowski:2014:JCTC}
Janowski,~T. Near Equivalence of Intrinsic Atomic Orbitals and Quasiatomic
  Orbitals. \emph{J. Chem. Theory Comput.} \textbf{2014}, \emph{10},
  3085--3091\relax
\mciteBstWouldAddEndPuncttrue
\mciteSetBstMidEndSepPunct{\mcitedefaultmidpunct}
{\mcitedefaultendpunct}{\mcitedefaultseppunct}\relax
\EndOfBibitem
\bibitem[H{\o}yvik \latin{et~al.}(2012)H{\o}yvik, Jansik, and
  J{\o}rgensen]{VRG:hoyvik:2012:JCP}
H{\o}yvik,~I.-M.; Jansik,~B.; J{\o}rgensen,~P. Orbital Localization Using
  Fourth Central Moment Minimization. \emph{J. Chem. Phys.} \textbf{2012},
  \emph{137}, 224114\relax
\mciteBstWouldAddEndPuncttrue
\mciteSetBstMidEndSepPunct{\mcitedefaultmidpunct}
{\mcitedefaultendpunct}{\mcitedefaultseppunct}\relax
\EndOfBibitem
\bibitem[Marzari and Vanderbilt(1997)Marzari, and
  Vanderbilt]{VRG:marzari:1997:PRB}
Marzari,~N.; Vanderbilt,~D. Maximally Localized Generalized {{Wannier}}
  Functions for Composite Energy Bands. \emph{Phys. Rev. B} \textbf{1997},
  \emph{56}, 12847--12865\relax
\mciteBstWouldAddEndPuncttrue
\mciteSetBstMidEndSepPunct{\mcitedefaultmidpunct}
{\mcitedefaultendpunct}{\mcitedefaultseppunct}\relax
\EndOfBibitem
\bibitem[Wannier(1937)]{VRG:wannier:1937:PR}
Wannier,~G.~H. The {{Structure}} of {{Electronic Excitation Levels}} in
  {{Insulating Crystals}}. \emph{Phys. Rev.} \textbf{1937}, \emph{52},
  191--197\relax
\mciteBstWouldAddEndPuncttrue
\mciteSetBstMidEndSepPunct{\mcitedefaultmidpunct}
{\mcitedefaultendpunct}{\mcitedefaultseppunct}\relax
\EndOfBibitem
\bibitem[Marzari \latin{et~al.}(2012)Marzari, Mostofi, Yates, Souza, and
  Vanderbilt]{VRG:marzari:2012:RMP}
Marzari,~N.; Mostofi,~A.~A.; Yates,~J.~R.; Souza,~I.; Vanderbilt,~D. Maximally
  Localized {{Wannier}} Functions: {{Theory}} and Applications. \emph{Rev. Mod.
  Phys.} \textbf{2012}, \emph{84}, 1419--1475\relax
\mciteBstWouldAddEndPuncttrue
\mciteSetBstMidEndSepPunct{\mcitedefaultmidpunct}
{\mcitedefaultendpunct}{\mcitedefaultseppunct}\relax
\EndOfBibitem
\bibitem[Pizzi \latin{et~al.}(2020)Pizzi, Vitale, Arita, Bl{\"u}gel, Freimuth,
  G{\'e}ranton, Gibertini, Gresch, Johnson, Koretsune, {Iba{\~n}ez-Azpiroz},
  Lee, Lihm, Marchand, Marrazzo, Mokrousov, Mustafa, Nohara, Nomura, Paulatto,
  Ponc{\'e}, Ponweiser, Qiao, Th{\"o}le, Tsirkin, Wierzbowska, Marzari,
  Vanderbilt, Souza, Mostofi, and Yates]{VRG:pizzi:2020:JPCM}
Pizzi,~G.; Vitale,~V.; Arita,~R.; Bl{\"u}gel,~S.; Freimuth,~F.;
  G{\'e}ranton,~G.; Gibertini,~M.; Gresch,~D.; Johnson,~C.; Koretsune,~T.;
  {Iba{\~n}ez-Azpiroz},~J.; Lee,~H.; Lihm,~J.-M.; Marchand,~D.; Marrazzo,~A.;
  Mokrousov,~Y.; Mustafa,~J.~I.; Nohara,~Y.; Nomura,~Y.; Paulatto,~L.;
  Ponc{\'e},~S.; Ponweiser,~T.; Qiao,~J.; Th{\"o}le,~F.; Tsirkin,~S.~S.;
  Wierzbowska,~M.; Marzari,~N.; Vanderbilt,~D.; Souza,~I.; Mostofi,~A.~A.;
  Yates,~J.~R. Wannier90 as a Community Code: New Features and Applications.
  \emph{J. Phys.: Condens. Matter} \textbf{2020}, \emph{32}, 165902\relax
\mciteBstWouldAddEndPuncttrue
\mciteSetBstMidEndSepPunct{\mcitedefaultmidpunct}
{\mcitedefaultendpunct}{\mcitedefaultseppunct}\relax
\EndOfBibitem
\bibitem[Maschio \latin{et~al.}(2007)Maschio, Usvyat, Manby, Casassa, Pisani,
  and Sch{\"u}tz]{VRG:maschio:2007:PRB}
Maschio,~L.; Usvyat,~D.; Manby,~F.~R.; Casassa,~S.; Pisani,~C.; Sch{\"u}tz,~M.
  Fast Local-{{MP2}} Method with Density-Fitting for Crystals. {{I}}.
  {{Theory}} and Algorithms. \emph{Phys. Rev. B} \textbf{2007}, \emph{76},
  8772\relax
\mciteBstWouldAddEndPuncttrue
\mciteSetBstMidEndSepPunct{\mcitedefaultmidpunct}
{\mcitedefaultendpunct}{\mcitedefaultseppunct}\relax
\EndOfBibitem
\bibitem[Usvyat \latin{et~al.}(2011)Usvyat, Civalleri, Maschio, Dovesi, Pisani,
  and Sch{\"u}tz]{VRG:usvyat:2011:JCP}
Usvyat,~D.; Civalleri,~B.; Maschio,~L.; Dovesi,~R.; Pisani,~C.; Sch{\"u}tz,~M.
  Approaching the Theoretical Limit in Periodic Local {{MP2}} Calculations with
  Atomic-Orbital Basis Sets: {{The}} Case of {{LiH}}. \emph{J. Chem. Phys.}
  \textbf{2011}, \emph{134}, 214105\relax
\mciteBstWouldAddEndPuncttrue
\mciteSetBstMidEndSepPunct{\mcitedefaultmidpunct}
{\mcitedefaultendpunct}{\mcitedefaultseppunct}\relax
\EndOfBibitem
\bibitem[{Zicovich-Wilson} \latin{et~al.}(2001){Zicovich-Wilson}, Dovesi, and
  Saunders]{VRG:zicovich-wilson:2001:JCP}
{Zicovich-Wilson},~C.~M.; Dovesi,~R.; Saunders,~V.~R. A General Method to
  Obtain Well Localized {{Wannier}} Functions for Composite Energy Bands in
  Linear Combination of Atomic Orbital Periodic Calculations. \emph{J. Chem.
  Phys.} \textbf{2001}, \emph{115}, 9708--9719\relax
\mciteBstWouldAddEndPuncttrue
\mciteSetBstMidEndSepPunct{\mcitedefaultmidpunct}
{\mcitedefaultendpunct}{\mcitedefaultseppunct}\relax
\EndOfBibitem
\bibitem[Dovesi \latin{et~al.}(2018)Dovesi, Erba, Orlando, {Zicovich-Wilson},
  Civalleri, Maschio, R{\'e}rat, Casassa, Baima, Salustro, and
  Kirtman]{VRG:dovesi:2018:WIRCMS}
Dovesi,~R.; Erba,~A.; Orlando,~R.; {Zicovich-Wilson},~C.~M.; Civalleri,~B.;
  Maschio,~L.; R{\'e}rat,~M.; Casassa,~S.; Baima,~J.; Salustro,~S.; Kirtman,~B.
  Quantum-Mechanical Condensed Matter Simulations with {{CRYSTAL}}. \emph{Wiley
  Interdiscip. Rev. Comput. Mol. Sci.} \textbf{2018}, \emph{8}, e1360\relax
\mciteBstWouldAddEndPuncttrue
\mciteSetBstMidEndSepPunct{\mcitedefaultmidpunct}
{\mcitedefaultendpunct}{\mcitedefaultseppunct}\relax
\EndOfBibitem
\bibitem[J{\'o}nsson \latin{et~al.}(2017)J{\'o}nsson, Lehtola, Puska, and
  J{\'o}nsson]{VRG:jonsson:2017:JCTC}
J{\'o}nsson,~E.~{\"O}.; Lehtola,~S.; Puska,~M.; J{\'o}nsson,~H. Theory and
  {{Applications}} of {{Generalized Pipek}}\textendash{{Mezey Wannier
  Functions}}. \emph{J. Chem. Theory Comput.} \textbf{2017}, \emph{13},
  460--474\relax
\mciteBstWouldAddEndPuncttrue
\mciteSetBstMidEndSepPunct{\mcitedefaultmidpunct}
{\mcitedefaultendpunct}{\mcitedefaultseppunct}\relax
\EndOfBibitem
\bibitem[Lehtola and J{\'o}nsson(2013)Lehtola, and
  J{\'o}nsson]{VRG:lehtola:2013:JCTC}
Lehtola,~S.; J{\'o}nsson,~H. Unitary Optimization of Localized Molecular
  Orbitals. \emph{J. Chem. Theory Comput.} \textbf{2013}, \emph{9},
  5365--5372\relax
\mciteBstWouldAddEndPuncttrue
\mciteSetBstMidEndSepPunct{\mcitedefaultmidpunct}
{\mcitedefaultendpunct}{\mcitedefaultseppunct}\relax
\EndOfBibitem
\bibitem[Broyden(1970)]{VRG:broyden:1970:IJAM}
Broyden,~C.~G. The {{Convergence}} of a {{Class}} of {{Double}}-Rank
  {{Minimization Algorithms}} 1. {{General Considerations}}. \emph{IMA J.
  Applies Math.} \textbf{1970}, \emph{6}, 76--90\relax
\mciteBstWouldAddEndPuncttrue
\mciteSetBstMidEndSepPunct{\mcitedefaultmidpunct}
{\mcitedefaultendpunct}{\mcitedefaultseppunct}\relax
\EndOfBibitem
\bibitem[Fletcher(1970)]{VRG:fletcher:1970:CJ}
Fletcher,~R. A New Approach to Variable Metric Algorithms. \emph{Comput. J.}
  \textbf{1970}, \emph{13}, 317--322\relax
\mciteBstWouldAddEndPuncttrue
\mciteSetBstMidEndSepPunct{\mcitedefaultmidpunct}
{\mcitedefaultendpunct}{\mcitedefaultseppunct}\relax
\EndOfBibitem
\bibitem[Goldfarb(1970)]{VRG:goldfarb:1970:MC}
Goldfarb,~D. A {{Family}} of {{Variable}}-{{Metric Methods Derived}} by
  {{Variational Means}}. \emph{Math. Comp.} \textbf{1970}, \emph{24},
  23--26\relax
\mciteBstWouldAddEndPuncttrue
\mciteSetBstMidEndSepPunct{\mcitedefaultmidpunct}
{\mcitedefaultendpunct}{\mcitedefaultseppunct}\relax
\EndOfBibitem
\bibitem[Shanno(1970)]{VRG:shanno:1970:MC}
Shanno,~D.~F. Conditioning of Quasi-{{Newton}} Methods for Function
  Minimization. \emph{Math. Comp.} \textbf{1970}, \emph{24}, 647--647\relax
\mciteBstWouldAddEndPuncttrue
\mciteSetBstMidEndSepPunct{\mcitedefaultmidpunct}
{\mcitedefaultendpunct}{\mcitedefaultseppunct}\relax
\EndOfBibitem
\bibitem[Kari(1984)]{VRG:kari:1984:IJQC}
Kari,~R. Parametrization and Comparative Analysis of {{theBFGS}} Optimization
  Algorithm for the Determination of Optimum Linear Coefficients. \emph{Int. J.
  Quantum Chem.} \textbf{1984}, \emph{25}, 321--329\relax
\mciteBstWouldAddEndPuncttrue
\mciteSetBstMidEndSepPunct{\mcitedefaultmidpunct}
{\mcitedefaultendpunct}{\mcitedefaultseppunct}\relax
\EndOfBibitem
\bibitem[Leonard and Luken(1982)Leonard, and Luken]{VRG:leonard:1982:TCA}
Leonard,~J.~M.; Luken,~W.~L. Quadratically Convergent Calculation of Localized
  Molecular Orbitals. \emph{Theoret. Chim. Acta} \textbf{1982}, \emph{62},
  107--132\relax
\mciteBstWouldAddEndPuncttrue
\mciteSetBstMidEndSepPunct{\mcitedefaultmidpunct}
{\mcitedefaultendpunct}{\mcitedefaultseppunct}\relax
\EndOfBibitem
\bibitem[H{\o}yvik \latin{et~al.}(2012)H{\o}yvik, Jansik, and
  J{\o}rgensen]{VRG:hoyvik:2012:JCTC}
H{\o}yvik,~I.-M.; Jansik,~B.; J{\o}rgensen,~P. Trust {{Region Minimization}} of
  {{Orbital Localization Functions}}. \emph{J. Chem. Theory Comput.}
  \textbf{2012}, \emph{8}, 3137--3146\relax
\mciteBstWouldAddEndPuncttrue
\mciteSetBstMidEndSepPunct{\mcitedefaultmidpunct}
{\mcitedefaultendpunct}{\mcitedefaultseppunct}\relax
\EndOfBibitem
\bibitem[Wang \latin{et~al.}(2020)Wang, Lewis, and Valeev]{VRG:wang:2020:JCP}
Wang,~X.; Lewis,~C.~A.; Valeev,~E.~F. Efficient Evaluation of Exact Exchange
  for Periodic Systems via Concentric Atomic Density Fitting. \emph{J. Chem.
  Phys.} \textbf{2020}, \emph{153}, 124116\relax
\mciteBstWouldAddEndPuncttrue
\mciteSetBstMidEndSepPunct{\mcitedefaultmidpunct}
{\mcitedefaultendpunct}{\mcitedefaultseppunct}\relax
\EndOfBibitem
\bibitem[Harris(1975)]{VRG:harris:1975:Tc}
Harris,~F.~E. In \emph{Theoretical Chemistry}; Eyring,~H., Henderson,~D., Eds.;
  {Elsevier}, 1975; pp 147--218\relax
\mciteBstWouldAddEndPuncttrue
\mciteSetBstMidEndSepPunct{\mcitedefaultmidpunct}
{\mcitedefaultendpunct}{\mcitedefaultseppunct}\relax
\EndOfBibitem
\bibitem[Pisani and Dovesi(1980)Pisani, and Dovesi]{VRG:pisani:1980:IJQC}
Pisani,~C.; Dovesi,~R. Exact-Exchange {{Hartree}}-{{Fock}} Calculations for
  Periodic Systems. {{I}}. {{Illustration}} of the Method. \emph{Int. J.
  Quantum Chem.} \textbf{1980}, \emph{17}, 501--516\relax
\mciteBstWouldAddEndPuncttrue
\mciteSetBstMidEndSepPunct{\mcitedefaultmidpunct}
{\mcitedefaultendpunct}{\mcitedefaultseppunct}\relax
\EndOfBibitem
\bibitem[Monkhorst and Pack(1976)Monkhorst, and Pack]{VRG:monkhorst:1976:PRB}
Monkhorst,~H.~J.; Pack,~J.~D. Special Points for {{Brillouin}}-Zone
  Integrations. \emph{Phys. Rev. B} \textbf{1976}, \emph{13}, 5188--5192\relax
\mciteBstWouldAddEndPuncttrue
\mciteSetBstMidEndSepPunct{\mcitedefaultmidpunct}
{\mcitedefaultendpunct}{\mcitedefaultseppunct}\relax
\EndOfBibitem
\bibitem[West \latin{et~al.}(2013)West, Schmidt, Gordon, and
  Ruedenberg]{VRG:west:2013:JCP}
West,~A.~C.; Schmidt,~M.~W.; Gordon,~M.~S.; Ruedenberg,~K. A Comprehensive
  Analysis of Molecule-Intrinsic Quasi-Atomic, Bonding, and Correlating
  Orbitals. {{I}}. {{Hartree}}-{{Fock}} Wave Functions. \emph{J. Chem. Phys.}
  \textbf{2013}, \emph{139}, 234107\relax
\mciteBstWouldAddEndPuncttrue
\mciteSetBstMidEndSepPunct{\mcitedefaultmidpunct}
{\mcitedefaultendpunct}{\mcitedefaultseppunct}\relax
\EndOfBibitem
\bibitem[Mulliken(1962)]{VRG:mulliken:1962:JCP}
Mulliken,~R.~S. Criteria for the {{Construction}} of {{Good
  Self}}-{{Consistent}}-{{Field Molecular Orbital Wave Functions}}, and the
  {{Significance}} of {{LCAO}}-{{MO Population Analysis}}. \emph{J. Chem.
  Phys.} \textbf{1962}, \emph{36}, 3428--3439\relax
\mciteBstWouldAddEndPuncttrue
\mciteSetBstMidEndSepPunct{\mcitedefaultmidpunct}
{\mcitedefaultendpunct}{\mcitedefaultseppunct}\relax
\EndOfBibitem
\bibitem[Lu \latin{et~al.}(2004)Lu, Wang, Schmidt, Bytautas, Ho, and
  Ruedenberg]{VRG:lu:2004:JCP}
Lu,~W.~C.; Wang,~C.~Z.; Schmidt,~M.~W.; Bytautas,~L.; Ho,~K.~M.; Ruedenberg,~K.
  Molecule Intrinsic Minimal Basis Sets. {{I}}. {{Exact}} Resolution of
  {\emph{Ab Initio}} Optimized Molecular Orbitals in Terms of Deformed Atomic
  Minimal-Basis Orbitals. \emph{J. Chem. Phys.} \textbf{2004}, \emph{120},
  2629--2637\relax
\mciteBstWouldAddEndPuncttrue
\mciteSetBstMidEndSepPunct{\mcitedefaultmidpunct}
{\mcitedefaultendpunct}{\mcitedefaultseppunct}\relax
\EndOfBibitem
\bibitem[Laikov(2011)]{VRG:laikov:2011:IJQC}
Laikov,~D.~N. Intrinsic Minimal Atomic Basis Representation of Molecular
  Electronic Wavefunctions. \emph{Int. J. Quantum Chem.} \textbf{2011},
  \emph{111}, 2851--2867\relax
\mciteBstWouldAddEndPuncttrue
\mciteSetBstMidEndSepPunct{\mcitedefaultmidpunct}
{\mcitedefaultendpunct}{\mcitedefaultseppunct}\relax
\EndOfBibitem
\bibitem[King \latin{et~al.}(1967)King, Stanton, Kim, Wyatt, and
  Parr]{VRG:king:1967:JCP}
King,~H.~F.; Stanton,~R.~E.; Kim,~H.; Wyatt,~R.~E.; Parr,~R.~G. Corresponding
  {{Orbitals}} and the {{Nonorthogonality Problem}} in {{Molecular Quantum
  Mechanics}}. \emph{J. Chem. Phys.} \textbf{1967}, \emph{47}, 1936--1941\relax
\mciteBstWouldAddEndPuncttrue
\mciteSetBstMidEndSepPunct{\mcitedefaultmidpunct}
{\mcitedefaultendpunct}{\mcitedefaultseppunct}\relax
\EndOfBibitem
\bibitem[Mustafa \latin{et~al.}(2015)Mustafa, Coh, Cohen, and
  Louie]{VRG:mustafa:2015:PRB}
Mustafa,~J.~I.; Coh,~S.; Cohen,~M.~L.; Louie,~S.~G. Automated Construction of
  Maximally Localized {{Wannier}} Functions: {{Optimized}} Projection Functions
  Method. \emph{Phys. Rev. B} \textbf{2015}, \emph{92}, 165134\relax
\mciteBstWouldAddEndPuncttrue
\mciteSetBstMidEndSepPunct{\mcitedefaultmidpunct}
{\mcitedefaultendpunct}{\mcitedefaultseppunct}\relax
\EndOfBibitem
\bibitem[{Brandwood}(1983)]{Brandwood1983}
{Brandwood},~D.~H. {A complex gradient operator and its application in adaptive
  array theory}. \emph{IEE Proceedings F: Communications Radar and Signal
  Processing} \textbf{1983}, \emph{130}, 11--16\relax
\mciteBstWouldAddEndPuncttrue
\mciteSetBstMidEndSepPunct{\mcitedefaultmidpunct}
{\mcitedefaultendpunct}{\mcitedefaultseppunct}\relax
\EndOfBibitem
\bibitem[Abrudan \latin{et~al.}(2008)Abrudan, Eriksson, and
  Koivunen]{VRG:abrudan:2008:ITSP}
Abrudan,~T.~E.; Eriksson,~J.; Koivunen,~V. Steepest Descent Algorithms for
  Optimization under Unitary Matrix Constraint. \emph{IEEE Trans. Signal
  Process.} \textbf{2008}, \emph{56}, 1134--1147\relax
\mciteBstWouldAddEndPuncttrue
\mciteSetBstMidEndSepPunct{\mcitedefaultmidpunct}
{\mcitedefaultendpunct}{\mcitedefaultseppunct}\relax
\EndOfBibitem
\bibitem[Abrudan \latin{et~al.}(2009)Abrudan, Eriksson, and Koivunen]{abrudan}
Abrudan,~T.; Eriksson,~J.; Koivunen,~V. Conjugate gradient algorithm for
  optimization under unitary matrix constraint. \emph{Signal Processing}
  \textbf{2009}, \emph{89}, 1704--1714\relax
\mciteBstWouldAddEndPuncttrue
\mciteSetBstMidEndSepPunct{\mcitedefaultmidpunct}
{\mcitedefaultendpunct}{\mcitedefaultseppunct}\relax
\EndOfBibitem
\bibitem[Polak and Ribi\`{e}re(1969)Polak, and Ribi\`{e}re]{cgpr}
Polak,~E.; Ribi\`{e}re,~G. Note sur la convergence de m\'ethodes de directions
  conjugu\'ees. \emph{ESAIM: Mathematical Modelling and Numerical Analysis -
  Mod\'elisation Math\'ematique et Analyse Num\'erique} \textbf{1969},
  \emph{3}, 35--43\relax
\mciteBstWouldAddEndPuncttrue
\mciteSetBstMidEndSepPunct{\mcitedefaultmidpunct}
{\mcitedefaultendpunct}{\mcitedefaultseppunct}\relax
\EndOfBibitem
\bibitem[Fletcher and Reeves(1964)Fletcher, and Reeves]{cgfr}
Fletcher,~R.; Reeves,~C.~M. Function minimization by conjugate gradients.
  \emph{The Computer Journal} \textbf{1964}, \emph{7}, 149--154\relax
\mciteBstWouldAddEndPuncttrue
\mciteSetBstMidEndSepPunct{\mcitedefaultmidpunct}
{\mcitedefaultendpunct}{\mcitedefaultseppunct}\relax
\EndOfBibitem
\bibitem[Hestenes and Stiefel(1952)Hestenes, and Stiefel]{cghs}
Hestenes,~M.~R.; Stiefel,~E. Methods of conjugate gradients for solving linear
  systems. \emph{Journal of research of the National Bureau of Standards}
  \textbf{1952}, \emph{49}, 409--435\relax
\mciteBstWouldAddEndPuncttrue
\mciteSetBstMidEndSepPunct{\mcitedefaultmidpunct}
{\mcitedefaultendpunct}{\mcitedefaultseppunct}\relax
\EndOfBibitem
\bibitem[Sorber \latin{et~al.}(2012)Sorber, Barel, and
  Lathauwer]{VRG:sorber:2012:SJO}
Sorber,~L.; Barel,~M.~V.; Lathauwer,~L.~D. Unconstrained {{Optimization}} of
  {{Real Functions}} in {{Complex Variables}}. \emph{SIAM J. Optim.}
  \textbf{2012}, \emph{22}, 879--898\relax
\mciteBstWouldAddEndPuncttrue
\mciteSetBstMidEndSepPunct{\mcitedefaultmidpunct}
{\mcitedefaultendpunct}{\mcitedefaultseppunct}\relax
\EndOfBibitem
\bibitem[Nocedal(1980)]{VRG:nocedal:1980:MC}
Nocedal,~J. Updating Quasi-{{Newton}} Matrices with Limited Storage.
  \emph{Math. Comp.} \textbf{1980}, \emph{35}, 773--773\relax
\mciteBstWouldAddEndPuncttrue
\mciteSetBstMidEndSepPunct{\mcitedefaultmidpunct}
{\mcitedefaultendpunct}{\mcitedefaultseppunct}\relax
\EndOfBibitem
\bibitem[Peng \latin{et~al.}(2020)Peng, Lewis, Wang, Clement, Pierce, Rishi,
  Pavo\u{s}evi\'{c}, Slattery, Zhang, Teke, Kumar, Masteran, Asadchev, Calvin,
  and Valeev]{mpqc}
Peng,~C.; Lewis,~C.~A.; Wang,~X.; Clement,~M.~C.; Pierce,~K.; Rishi,~V.;
  Pavo\u{s}evi\'{c},~F.; Slattery,~S.; Zhang,~J.; Teke,~N.; Kumar,~A.;
  Masteran,~C.; Asadchev,~A.; Calvin,~J.~A.; Valeev,~E.~F. {M}assively
  {P}arallel {Q}uantum {C}hemistry: {A} high-performance research platform for
  electronic structure. \emph{J. Chem. Phys.} \textbf{2020}, \emph{153},
  044120\relax
\mciteBstWouldAddEndPuncttrue
\mciteSetBstMidEndSepPunct{\mcitedefaultmidpunct}
{\mcitedefaultendpunct}{\mcitedefaultseppunct}\relax
\EndOfBibitem
\bibitem[VTA()]{VTARCNewRiver}
Newriver compute cluster at Virginia Tech.
  \url{https://www.arc.vt.edu/computing/newriver} (accessed Jan 26,
  2021).\relax
\mciteBstWouldAddEndPunctfalse
\mciteSetBstMidEndSepPunct{\mcitedefaultmidpunct}
{}{\mcitedefaultseppunct}\relax
\EndOfBibitem
\bibitem[Weigend(2006)]{weigend2006}
Weigend,~F. Accurate Coulomb-fitting basis sets for H to Rn. \emph{Phys. Chem.
  Chem. Phys.} \textbf{2006}, \emph{8}, 1057 -- 1065\relax
\mciteBstWouldAddEndPuncttrue
\mciteSetBstMidEndSepPunct{\mcitedefaultmidpunct}
{\mcitedefaultendpunct}{\mcitedefaultseppunct}\relax
\EndOfBibitem
\bibitem[van Duijneveldt(1971)]{MINI:H}
van Duijneveldt,~F.~B. \emph{Gaussian basis sets for the atoms H-Ne for use in
  molecular calculations}; 1971\relax
\mciteBstWouldAddEndPuncttrue
\mciteSetBstMidEndSepPunct{\mcitedefaultmidpunct}
{\mcitedefaultendpunct}{\mcitedefaultseppunct}\relax
\EndOfBibitem
\bibitem[Andzelm \latin{et~al.}(1984)Andzelm, Huzinaga, Klobukowski,
  Radzio-Andzelm, Sakai, and Tatewaki]{MINI}
Andzelm,~J.; Huzinaga,~S.; Klobukowski,~M.; Radzio-Andzelm,~E.; Sakai,~Y.;
  Tatewaki,~H. In \emph{Gaussian Basis Sets for Molecular Calculations};
  Huzinaga,~S., Ed.; Physical Sciences Data; Elsevier, 1984; Vol.~16; Chapter
  Gaussian Basis Sets, pp 27--426\relax
\mciteBstWouldAddEndPuncttrue
\mciteSetBstMidEndSepPunct{\mcitedefaultmidpunct}
{\mcitedefaultendpunct}{\mcitedefaultseppunct}\relax
\EndOfBibitem
\bibitem[Schuchardt \latin{et~al.}(2007)Schuchardt, Didier, Elsethagen, Sun,
  Gurumoorthi, Chase, Li, and Windus]{BSE}
Schuchardt,~K.~L.; Didier,~B.~T.; Elsethagen,~T.; Sun,~L.; Gurumoorthi,~V.;
  Chase,~J.; Li,~J.; Windus,~T.~L. Basis Set Exchange: A Community Database for
  Computational Sciences. \emph{J. Chem. Inf. Model.} \textbf{2007}, \emph{47},
  1045--1052\relax
\mciteBstWouldAddEndPuncttrue
\mciteSetBstMidEndSepPunct{\mcitedefaultmidpunct}
{\mcitedefaultendpunct}{\mcitedefaultseppunct}\relax
\EndOfBibitem
\bibitem[Lorenz \latin{et~al.}(2012)Lorenz, Maschio, Sch{\"u}tz, and
  Usvyat]{VRG:lorenz:2012:JCP}
Lorenz,~M.; Maschio,~L.; Sch{\"u}tz,~M.; Usvyat,~D. Local Ab Initio Methods for
  Calculating Optical Bandgaps in Periodic Systems. {{II}}. {{Periodic}}
  Density Fitted Local Configuration Interaction Singles Method for Solids.
  \emph{J. Chem. Phys.} \textbf{2012}, \emph{137}, 204119\relax
\mciteBstWouldAddEndPuncttrue
\mciteSetBstMidEndSepPunct{\mcitedefaultmidpunct}
{\mcitedefaultendpunct}{\mcitedefaultseppunct}\relax
\EndOfBibitem
\bibitem[Axelsson(2003)]{VRG:axelsson:2003:MaCiSa}
Axelsson,~O. Iteration Number for the Conjugate Gradient Method.
  \emph{Mathematics and Computers in Simulation} \textbf{2003}, \emph{61},
  421--435\relax
\mciteBstWouldAddEndPuncttrue
\mciteSetBstMidEndSepPunct{\mcitedefaultmidpunct}
{\mcitedefaultendpunct}{\mcitedefaultseppunct}\relax
\EndOfBibitem
\bibitem[Lehtola \latin{et~al.}(2016)Lehtola, {Head-Gordon}, and
  J{\'o}nsson]{VRG:lehtola:2016:JCTC}
Lehtola,~S.; {Head-Gordon},~M.; J{\'o}nsson,~H. Complex Orbitals, Multiple
  Local Minima, and Symmetry Breaking in {{Perdew}}\textendash{{Zunger}}
  Self-Interaction Corrected Density Functional Theory Calculations. \emph{J.
  Chem. Theory Comput.} \textbf{2016}, \emph{12}, 3195\relax
\mciteBstWouldAddEndPuncttrue
\mciteSetBstMidEndSepPunct{\mcitedefaultmidpunct}
{\mcitedefaultendpunct}{\mcitedefaultseppunct}\relax
\EndOfBibitem
\end{mcitethebibliography}

\end{document}